\documentclass[nofootinbib,eqsecnum,tightenlines,10pt]{revtex4}

\usepackage{graphicx}
\usepackage{amsmath,amssymb,amsfonts,amsthm,stmaryrd,mathtools,bm}
\usepackage{mathrsfs}
\usepackage{color}
\usepackage{multirow,bigdelim}
\usepackage{hyperref}
\usepackage{dsfont}
\usepackage{booktabs}
\allowdisplaybreaks[0]

\usepackage{pstricks}
\usepackage{tikz}
\usepackage{ulem}

\def\be#1\ee{\begin{align}#1\end{align}}

\def\ba{\begin{eqnarray}}
\def\ea{\end{eqnarray}}
\def\nn{\nonumber}
\def\q{\quad}

\begin{document}

\title{  Modified Graviton Dynamics From  Spin Foams: The Area Regge Action}

\author{Bianca Dittrich}
\affiliation{Perimeter Institute, 31 Caroline Street North, Waterloo, ON, N2L 2Y5, CAN}

\begin{abstract}
A number of approaches to 4D quantum gravity, such as holography and loop quantum gravity, propose areas instead of lengths as fundamental variables. The Area Regge action, which can be defined for general 4D triangulations, is a natural choice for an action based on areas.  It does indeed appear in the semi-classical limit of spin foam models. The Area Regge  action does however only lead to a discrete version of the gravitational equations of motion, if one implements constraints, that ensure that the areas are compatible with a consistent length assignment to the edges of the triangulation. The constrained version is then classically equivalent to the Length Regge action, which provides a discretization of the Einstein-Hilbert action.
 
Here we perform the first systematic analysis of the Area Regge dynamics on a hyper-cubical lattice. Surprisingly, we find that the linearized Area Regge action on a hyper-cubical lattice does single out the Length Regge action  by its scaling behaviour in the lattice constant. That is, integrating out the variables describing fluctuations in the area-length constraints one finds the linearized Length Regge action plus terms of higher order in the lattice constant. This appears without any explicit  implementation of the area-length constraints.

\end{abstract}

\maketitle

\section{Introduction}

Many approaches to quantum gravity are based on  a notion of quantum geometry. Here one faces the important question, which geometric variables are providing the fundamental degrees of freedom. Areas appear as natural choice for four-dimensional quantum gravity in e.g.  loop quantum gravity  and spin foams \cite{LQG,Perez} or holography \cite{RyuTakayanagi}.

Spin foam models provide a path integral for quantum geometries, which are described by triangulations, where the triangles carry area variables.\footnote{The more recent models also integrate over 3D angle degrees of freedom. Integrating these out gives an `effective' model, which sums over area variables only.}  The Area Regge action \cite{AreaRegge,ADH1} does appear in the (semi-) classical limit for spin foams \cite{SFLimit}.  This  action prescribes a dynamics for the discrete  geometries. Its equation of motion demand however that the (discrete) curvature is vanishing and thus do not seem to give a discretization of the equations of motion of general relativity. 

Associating area variables to the triangles of a four-dimensional triangulation, gives in general a much larger configuration space than the one obtained from associating length variables to the edges of a triangulation. But one can specify a set of constraints, that reduce the area configurations to length configurations. We will refer to these constraints as area-length constraints. 

The status of the area-length constraints in the various spin foam models \cite{ILQGS} has not been fully established yet. There are two exceptions:  firstly the Barrett-Crane model \cite{BC}, which was proposed first, features a locality structure that excludes the possibility that the constraints, which break this locality structure,  are implemented. Secondly, the recently introduced effective spin foam models \cite{EffSF1,EffSF2,EffSF3}, which do encode the dynamics in a more transparent way than previous models. These models do, in particular, include an explicit implementation of the area-length constraints. The area-length constraints feature an anomaly in their algebra \cite{DittrichRyan}, which  is parametrized by the Barbero--Immirzi parameter \cite{BarberoImmirzi} and only allows for a weak implementation \cite{EPRL-FK}. The effective spin foam models do therefore allow for fluctuations of the constraints, but with minimal uncertainty.  

This method of weakly imposing constraints has not been thoroughly tested. The work \cite{EffSF2} provided a first explicit test for the discrete equations of motion: Using a triangulation with only a few building blocks \cite{EffSF2} computed expectation values of geometric variables and identified a regime for the anomaly (Barbero--Immirzi) parameter, in which these expectation values reproduced well the classical solutions of Length Regge calculus. 

Having shown that the effective spin foams do admit a semi-classical limit for a coarse triangulation, the next pressing question is to understand the continuum limit. Here we mean with continuum limit the limit where the average lattice lengths is much smaller than the wave length of gravitational excitations, that we want to model. We have thus to consider lattices with many building blocks. Doing so we do not need to take the lattice constant to zero, and can even assume that it is of the order of $10^1$ or $10^2$ Planck lengths, as long as the wave lengths of the gravitational excitations we consider are much larger. The spin foam amplitudes show a good semi-classical behaviour\footnote{A lattice constant of $10^1$ to $10^2$ Planck lengths would translate into  background spin values of order $10^2$ to $10^4$. For Euclidean signature the spin foam amplitudes are well approximated by their semi-classical limit starting with spins of order $10^1$.} for such a range of the lattice constant.

The direct and non-perturbative evaluation of the  path integral is for such lattices with many building blocks  currently out of reach. But a first test for the continuum limit can be performed, by using a perturbative expansion (around flat space) of the action defining the effective spin foam models.   This action has two parts: a real part, given by the Area Regge action and an imaginary part, resulting from the weak implementation of the constraints. 

Expanding this action to second order in perturbations, and integrating out all area-length constraint degrees of freedom, we expect that we obtain the linearized Einstein Hilbert action plus correction terms, which result from the fluctuations of the area-length constraint degrees of freedom. The constraint terms in the action involve the Barbero--Immirzi parameter, one would therefore expect that this parameter does also parametrize the corrections.

The crucial question is to understand the scaling behaviour of these corrections in the lattice constant:  a scaling with negative powers would inform us that the continuum limit of spin foams does not lead to general relativity.  A scaling with positive powers of the lattice constant does indicate that spin foams can admit a suitable continuum limit. A derivation of these corrections would also help to understand what kind of quantum gravity effects we can expect from loop quantum gravity, and more generally from a theory in which areas provide the fundamental variables.

The perturbative expansion alone is of course not sufficient to proof that spin foams do admit a suitable continuum limit. For this one would need still a better understanding of the properties of the non-perturbative path integral, and in particular its divergence structure. But it is an important first step, and can already lead to phenomenological predications.

To start with we will  consider  in this work the Area Regge action without the constraint terms. Despite the important role of the Area Regge action for spin foams, there is not much work on understanding its kinematics and dynamics.\footnote{Previous works include \cite{Wainwright}, which aims to characterize the degrees of freedom for the Lorentzian signature case, and \cite{ADH1}, which provides a first-order formulation for Area Regge calculus (which resolves the problem with the ambiguities when solving for length in terms of areas) and an analysis of the propagating degrees of freedom  in local time evolution steps.}  The continuum limit is in particular not understood at all. One reason might be that Area Regge calculus appears to be problematic for lattices involving right angles, and in particular the hyper-cubical lattice. We will  show that this `problem' is rather helpful:  it happens to provide a mechanism which does suppress a certain part of the area-length constraint fluctuations. 

This mechanism might be essential for the rather surprising result we will find: the continuum limit of the linearized Area Regge action, on the hyper-cubical lattice  does lead to the linearized Einstein-Hilbert action. This occurs without any explicit implementation of the constraints. The lowest order correction term, resulting from integrating out the area-length degrees of freedom (which are not suppressed on the hyper-cubical lattice),  scales with the fourth power of the lattice constant, is quadratic in curvature, and is of sixth order in the momenta.\footnote{A perturbative expansion for spin foams was also suggested in \cite{Zipfel}. Using an additional coupling constant $\delta$ the authors added to the spin foam action  explicitly a term that suppresses the fluctuations in the area-length constraints. These constraints are not specified explicitly and their non-commutativity --- which  gives a bound on how much these fluctuations can be suppressed --- is not taking into account. Based on general arguments, that is without an explicit implementation of the linearized action on the hyper-cubic lattice,   \cite{Zipfel} finds that the correction term is proportional to $\delta^{-1}$ and argues that it is quadratic in curvature  and scales with $a_{\rm L}^{-2}$ where $a_{\rm L}$ is the lattice constant. Thus one assumes $\delta>>a_{\rm L}^{-2}$. Note however that $\delta$ cannot be chosen arbitrarily, the non-commutativity of the constraints do determine its scaling in $\gamma$ and the lattice constant. This scaling does depend on  the precise choice for the area-length constraints.
Our findings will differ in some essential ways:  here we do not add the area-constraints and find that the correction is nevertheless small, as it is of order $a_{\rm L}^4$  in the lattice constant. Based on a more thorough analysis of the Area Regge action, we find the explicit form of the correction, which is quadratic in first order derivatives of the curvature.}

This paper is structured as follows: We will start with a short over-view of Length and Area Regge calculus, and their role in (effective) spin foams, in Section \ref{Sec:Regge}. We discuss in Section \ref{Sec:Decoupling} the mechanism of decoupling or integrating out degrees of freedom. This mechanism is important to understand both the continuum limit of Length and Area Regge calculus. Section \ref{Sec:HC} constitutes the main part of the paper. It discusses the expansion of the Area and Length Regge action on the hyper-cubical lattice. This includes a discussion of the geometric interpretation of the degrees of freedom in Area Regge calculus as torsion degrees of freedom. A crucial part in understanding the properties of the Area Regge action is to find a convenient basis for the various types of degrees of freedom in Area Regge calculus, that we will encounter. This is discussed in Section \ref{Sec:separation}. The result for the continuum limit and the scaling of the lowest order correction, follows from the scaling of the various blocks of the Area Regge action Hessian, which are defined by the various types of degrees of freedom, see Section \ref{Sec:zeta}. We compute in Section \ref{Sec:Corr} the explicit form of the lowest order correction term.
We conclude with an outlook on including the constraint terms in the analysis in Section \ref{Outlook} and with a discussion in Section \ref{Disc}.

The expansion of the Area Regge action on the hyper-cubical lattice leads to rather involved expressions encoded in large matrices. We will therefore only  give few results, including the lowest order correction, in explicit form. But we will provide in Appendix \ref{App:Basis} a choice of basis for the various types of degrees of freedom, which we will encounter on the hyper-cubical lattice. This allows to reproduce the lowest order correction in the form presented here.

\section{Length and Area Regge calculus and (effective) spin foams}\label{Sec:Regge}

The Length Regge action \cite{Regge} can be defined for triangulations, which are decorated with a length variable  $L_e$ for each edge $e$. We will consider here four-dimensional triangulations, and work with geometries of Euclidean signature. The Length Regge action 
\ba\label{Eq:LR}
S_{\rm LR} =\frac{1}{\kappa} \sum_t A_t(L_e) \, \epsilon_t(L_e) \q ,
\ea
provides a discretization of the Einstein-Hilbert action.  $A_t(L_e)$ denotes the area of the triangle $t$ and 
\ba
\epsilon_t(L_e)=2\pi -\sum_{\sigma \supset t} \theta_t^\sigma(L_e)
\ea
is the deficit or curvature angle. $\theta_t^\sigma$ is the dihedral angle between two tetrahedra sharing the triangle $t$ in the four-simplex $\sigma$. Varying the Length Regge action with respect to the length variables we obtain the equations of motion\footnote{The Schl\"afli identity $\sum_{t\subset \sigma} A_t \delta \theta_t^\sigma=0$, which holds for arbitrary variations  $\delta$ of the geometry of the four-simplex $\sigma$, ensures that terms with derivatives of $\epsilon_t$ cancel out.}
\ba
\sum_t \frac{\partial A_t}{\partial L_e} \epsilon_t  = 0\q .
\ea
These equations of motion constitute a discretization of the vacuum Einstein equations, and thus admit curved solutions with $\epsilon_t\neq 0$. 

Choosing areas instead of lengths as independent variables  we can define a version of Regge calculus known as Area Regge calculus \cite{AreaRegge,ADH1}. Its action is given by
\ba
S_{\rm AR} =\frac{1}{\kappa} \sum_t A_t \, \epsilon_t(A_t)
\ea
where the deficit angle $\epsilon_t$ is now defined as
\ba
\epsilon_t(A_t) = 2\pi - \sum_{\sigma \supset t} \theta_t^\sigma( \{A_{t'}\}_{t'\subset \sigma}) \q .
\ea
The dihedral angles $\theta_t^\sigma$ of a four-simplex $\sigma$ can be computed from its lengths $L_e$. There are 10 edges  and therefore lengths $L_e$ for a four-simplex. These  also determine the  areas $A_t$ for the10 triangles of the simplex. We have the same number of lengths and areas, and can, locally\footnote{The ambiguities  in the area-length solutions are due to a choices of roots and can be resolved by using a first order formalism for Area Regge calculus \cite{ADH1} or by using an alternative formulation with areas and 3D angles \cite{AreaAngle}. We will however see that the singularities in the area-length solution system are crucial for the discussion in this paper, and due have physical consequences. These consequences also persist for the first order formulation and the area-angle formulation of Area Regge calculus, as both versions reduce (locally in configuration space) to Area Regge calculus after integrating out all other variables.} in configuration space, invert the lengths for the areas, that is define functions $L_e^\sigma(A_t)$. The dihedral angles as functions of the areas are then defined as $\theta_t^\sigma( \{A_{t'}\}_{t'\subset \sigma})=\theta_t^\sigma( L_e^\sigma(A_t))$.

Varying the Area Regge action with respect to the areas we do now get equations of motion
\ba
\epsilon_t(A_t)=0 \q ,
\ea
which impose vanishing deficit angles. There are however propagating degrees of freedom: Using a canonical formalism applicable for triangulations \cite{DittrichHoehn1}, the work \cite{ADH1} provided a counting of the propagating degrees  of freedom in linearized Area Regge calculus for so-called tent moves. A tent move constitutes a local time evolution of a given vertex in a triangulated hypersurface, which does not change the triangulation of this hypersurface. For a vertex with $n\geq4$ adjacent edges there are $(n-4)$ propagating degrees of freedom in Length Regge calculus. The $-4$ accounts for a discrete version of diffeomorphism symmetry \cite{RocekWilliams,DiffReview08}. But for Area Regge calculus, we have to count the triangles adjacent to an $n$--valent vertex. There are  $(3n-6)$ such triangles.  Accounting for the diffeomorphism symmetry we obtain $(3n-10)$ propagating degrees of freedom. We  have thus generically much more propagating degrees of freedom in Area Regge calculus than in Length Regge calculus. 

Indeed one can transform the area variables into two separate sets: firstly the length variables and secondly variables, which we will refer to as area-length constraints.  To explain these area-length constraints, we consider the gluing of two four-simplices along a shared tetrahedron. 'Gluing' means that we have to identify the geometric data, that is either lengths or areas, associated to the tetrahedron, which come from the two four-simplices.  A tetrahedron has six edges, but only four triangles.  For the glued complex we have therefore 20-6=14 lengths but 20-4=16 area variables.  

The additional two variables for Area Regge calculus account for the possibility that the geometries for the shared tetrahedron, as induced from the two four-simplices, differ. Indeed, in Area Regge calculus we only identify four pairs of areas for the shared tetrahedron. These are not sufficient to fix the complete geometry -- we rather need to also fix two additional geometric quantities, e.g. two length $(L_1,L_2)$ or two 3D dihedral angles $(\phi^\tau_1,\phi^\tau_2)$ in the tetrahedron. 

Consider a tetrahedron $\tau$, which is shared by four-simplices $\sigma$ and $\sigma'$. We can then determine the two additional edge lengths $(L^{\tau,\sigma}_1,L^{\tau,\sigma}_2)$ from the areas of $\sigma$ and $(L^{\tau,\sigma'}_1,L^{\tau,\sigma'}_2)$ from the areas of $\sigma'$. For a general area configuration we will have $L^{\tau,\sigma}_i \neq L^{\tau,\sigma'}_i$. If this is the case, the areas do not derive from a consistent assignment of lengths to the edges. We can therefore understand the differences $C^\tau_i= L^{\tau,\sigma}_i-L^{\tau,\sigma'}_i$ as constraints, whose vanishing ensures that the areas define a consistent length configuration. We will call (all possible versions of) these constraints area-length constraints.

Varying the Area Regge action only along directions tangent to the constraint hypersurface we regain the Length Regge equations of motion.\footnote{See \cite{EffSF3} for an explicit proof.}  To obtain a quantum gravity theory with general relativity as classical limit, it therefore seems to be necessary to add these constraints in some form to the Area Regge action. 

The Area Regge action is central in the spin foam approach \cite{Perez} to quantum gravity.  It arises in the large spin limit, which for a fixed discretization,  can be considered to be equivalent to the semi-classical $\hbar \rightarrow 0$ limit of spin foam amplitudes \cite{SFLimit}. The Area Regge action arose first for the Barrett-Crane model \cite{BC}, which can be considered to be a quantization of the Area Regge action \cite{AreaRegge,EffSF1}. In particular, the Barrett-Crane model does not implement the area-length constraints. The reason is the following: In contrast to later models \cite{EPRL-FK,BO} the Barrett-Crane model does only include the areas as fundamental variables. Its amplitude has also a specific locality property: it factorizes over the four-simplices (and triangles). Equivalently, there are two contributions to the Area Regge action, which can be split into two terms, one of which is a sum over triangles and the other a sum  over four-simplices:
\ba
S_{\rm AR} = \sum_t 2\pi A_t - \sum_\sigma \sum_{t\subset \sigma} \theta^\sigma_t( \{A_t\}_{t\in \sigma})  \q .
\ea
Importantly, the coefficient associated to a given triangle or four-simplex, can be determined from the data associated to this triangle or four-simplex alone.  The area-length constraints do break this locality structure: these are functions which do depend on the areas  belonging to pairs of neighbouring simplices. 
 Therefore, the Barrett-Crane model cannot implement the area-length constraints. 
 
 The more recent EPRL-FK spin foam models \cite{EPRL-FK} feature  also a version of the Regge action in their semi-classical limit. The EPRL-FK spin foam amplitudes do also factorize over four-simplices and triangles. These models do however include, apart from the areas, also quantum numbers that encode the two 3D dihedral angles $(\phi^\tau_1,\phi^\tau_2)$ per tetrahedron. Together with the areas these allow to completely specify the tetrahedral geometry.
 
 This inspired the proposal of the Area-Angle Regge action in \cite{AreaAngle}. This work provided the first explicit form\footnote{Other forms for the constraints have been used in \cite{DittrichRyan,EffSF1,EffSF3}.} for the area-length constraints (which were called gluing constraints\footnote{The constraints are also known as edge simplicity constraints \cite{DittrichRyan,BFCG2} or shape matching constraints \cite{Twisted}.}  in \cite{AreaAngle}). It also showed, that the 3D angles can be understood as auxiliary variables, which allow to implement the constraints without breaking the locality structure mentioned above. But integrating out the 3D angles results in an `effective' action for the areas only, which does break the locality structure, and does implement the area-length constraints.
 
 The EPRL-FK model might therefore be able to implement the area-length constraints. But whether this is the case, has not been established so far \cite{flatness}. The main reason for this is that the area-length constraints, are, due to an anomaly, of second class \cite{DittrichRyan, EffSF1}. More precisely, the two constraints $(C_1^\tau,C_2^\tau)$  associated to a given tetrahedron $\tau$ do not commute. This anomaly is parametrized by the Barbero-Immirzi parameter $\gamma$ \cite{BarberoImmirzi}. The constraints can therefore not be implemented exactly as this would violate the uncertainty principle. 
 
An alternative is to implement the constraints weakly.\footnote{This can be understood as a generalized Gupta-Bleuler treatment for the constraints.} The EPRL-FK models do apply such a weak implementation for the so-called (primary) simplicity constraints \cite{EPRL-FK}, that are also partially anomolous \cite{Perez}, with the anomaly parametrized by the Barbero-Immirzi parameter. In contrast, the area-length constraints arise from the secondary simplicity constraints \cite{DittrichRyan}. It has not been understood so far whether and how a weak implementation of primary constraints does lead to an implementation of secondary constraints. The status of the area-length constraints in the EPRL-FK models is therefore not clear. But they can at most be implemented weakly. 

The Barrett Crane model and the EPRL-FK model start with $\text{SO}(4)$ (or $\text{SO}(3,1)$) gauge variables. Areas and angles arise as gauge invariant quantities. The recently introduced effective spin foam model \cite{EffSF1,EffSF2,EffSF3} work directly with the areas and angles. This makes the encoding of the dynamics much more transparent, and the models far more accessible to numerical explorations. The action for these models is given by the Area Regge action plus an imaginary\footnote{The imaginary nature results from the constraints being implement by a Gaussian (measure).} term, that implements the Area-Length constraints weakly.  This term is quadratic in the constraints and its coefficient is fixed by the commutator between these constraints. Changing the constraints leads therefore to a change of the coefficients --- this makes this term covariant under a change of basis for the constraints at least to quadratic order in a perturbative expansion around a background. 

Clearly, understanding the dynamics of Area Regge calculus, and then the effect of implementing the constraints, will help to understand the dynamics encoded in spin foams. In this work we will concentrate on the dynamics of Area Regge calculus, defined on a hyper-cubic lattice. Astonishingly, it turns out, that the continuum limit of the linearized theory does lead to Length Regge calculus --- without any explicit implementation of the constraints.

%
%
%

\section{Integrating out vs. decoupling degrees of freedom}\label{Sec:Decoupling}

In many situations we can find a physically motivated splitting of the variables $\{x\}$ into two sets $\{y\}$ and $\{z\}$. For example in Area Regge calculus, we have, on the one hand, the length degrees of freedom, that is fluctuations of the areas  which result from the fluctuations of the length variables. On the other hand we do have also area fluctuations which violate the area--length constraints.   This distinction can be made independently of the dynamics. 

But it can also happen, that the dynamics distinguishes between certain sets of degrees of freedom. Here we will consider a linearized dynamics on the lattice, described by a quadratic action, which can be encoded into a (Hessian) matrix. We will find that certain sets of degrees of freedom are distinguished by the scaling behaviour of the corresponding matrix block in the lattice constant.  In such cases the continuum limit might require  to decouple this distinguished set of variables from the remaining ones.

To be more concrete, we consider a quadratic action 
\ba
S =\frac{1}{2}\left(y^\dagger  S_{yy} y  +  y^\dagger S_{yz} z + z^\dagger  S_{zy} y + z^\dagger  S_{zz}  z\right)  \q ,
\ea
where we assume a splitting of our variables into two sets $y\equiv \{y_a\}_{a=1}^{N_y}$ and $z\equiv \{z_b\}_{b=1}^{N_z}$, with $S_{yy}$ and $S_{zz}$ being symmetric (or Hermitian) matrices describing the coupling inbetween the $y$ and $z$ variables respectively, whereas for the  cross-diagonal blocks we have $S_{yz}=(S_{zy})^\dagger$.  Assuming that $S_{zz}$ is invertible, we define a variable transformation for the $z$-variables
\ba\label{dtrafo}
z=z'-S_{zz}^{-1}S_{zy} y \q .
\ea
In the new variables $(y,z')$ the action is given by
\ba
S=\frac{1}{2}\left(\, y^\dagger (S_{yy}- S_{yz} S_{zz}^{-1} S_{zy}) y   + (z')^\dagger S_{zz} z' \, \right) \q ,
\ea
with the $z'$ variables decoupled from the $y$ variables.

 If we are only interested in expectation values of the $y$--variables, we can consider 
 \ba
 S_y:=\tfrac{1}{2} y^\dagger(S_{yy}- S_{yz} S_{zz}^{-1} S_{zy}) y
 \ea
  as ``effective" action for the $y$--variables. One should however keep in mind that there is also an action $S_{z'}=(z')^\dagger S_{zz} z'$, and thus  equations of motion for the $z'$ variables.

This  decoupling method has been essential for the construction of the continuum limit for Length Regge calculus \cite{RocekWilliams}, and we will see that it plays an even more crucial role  for Area Regge calculus. 

We will encounter three types of situations, for which we will apply the decoupling transformation:

$(i)$ The first situation results from the fact that expanding the Length or Area Regge action in the lattice constant $a_L$, one will in general find that the various matrix elements are of in-homogeneous order in $a_L$. But one can identify a set of variables $z$, for which the eigenvalues of $S_{zz}$ scale homogeneously. Additionally $S_{zz}$ is local, that is does not involve lattice derivatives. Applying a decoupling transformation does lead to a homogeneous scaling for the $y$--block of the Hessian (or sub-blocks of the $y$--block). This  turns out to be  an important step for the construction of the continuum limit: In the cases discussed in this article the eigenvalues of $S_{zz}$ do diverge\footnote{This divergence behaviour could be changed by rescaling the variables with the lattice constant -- but this would not change the $a$-order of the correction term  $S_{yz} S_{zz}^{-1} S_{zy}$ in the lattice constant.} in the limit $a_L\rightarrow 0$. The $z'$--variables do then describe degrees of freedom which one has to scale to zero in order for the action to remain finite in the continuum limit. This mechanism was identified by Rocek and Williams in \cite{RocekWilliams} for Length Regge calculus. They considered the $z'$--variables as `spurious' variables, as they were not needed to reconstruct the metric perturbations from the length perturbations.  Barrett and Williams \cite{BarrettWilliams} performed an analysis based in independent arguments, that also showed that these spurious variables need to be scaled to zero, in order for the Regge solutions to approximate a continuum solution.

For a path integral the divergence of $S_{zz}$ in the $a_L\rightarrow 0$ limit means that the variances for the $z'$--degrees of freedom are vanishing in this limit. This is similar to these degrees of freedom becoming very heavy.

$(ii)$ In the second situation the various blocks $S_{yy}, S_{zz}$ and $S_{yz}$ have separately a homogeneous scaling, but  with  different powers of the lattice constant. This does appear for Area Regge calculus (after decoupling certain spurious variables), for a splitting of the variables into a set $\{y\}$ describing (length) metric fluctuations and a set $\{z\}$ describing fluctuations in (a certain subset of) the area-length constraints.  Specifically, we will find that $S_{yy}\sim a_L^0$  whereas $S_{zz}\sim a_L^{-2}$ and  $S_{yz}\sim a_L^1$.  Here one can already argue that only the $S_{yy}$ term -- which turns out to describe a lattice discretization of the linearized Einstein--Hilbert action -- does survive in the continuum limit. Indeed, integrating out the $z$-variables we obtain a correction term for the $y$--block, which with the scaling at hand, is of order $S_{yy}^{\rm Corr}\sim a^4_L$.

In short, we will find that the continuum limit for the Area Regge action on the hyper-cubic lattice does single out the Length Regge action part. This might be quite surprising, as the equations of motion of the Area Regge action impose a vanishing curvature, whereas the equations of motion of Length Regge calculus do only demand that certain contractions of deficit angles vanish, but does allow for curvature excitations. This can however be explained by looking at the scaling behaviour of the different parts in the equations of motion:
\ba
0&=&S_{yy} y - S_{yz} S_{zz}^{-1} S_{zy} y  \;\sim\, a_L^0 y - a_L^4 y  \label{3.4a}\\
0&=& S_{zz}z+  S_{zy}y    \; \;\q\q\q\sim\,       a_L^{-2} z + a_L^{1} y                    \label{3.4b}  
\ea
Taking together, Equations (\ref{3.4a}) and (\ref{3.4b}) do impose vanishing (linearized) deficit angles. The metric variables and the area-length constraints do appear however with different scalings. This forces an imposition of the area length constraints $z$ in the continuuum limit as  $z\sim a_L^3 y$. The equations for the metric fluctuations have also each two contributions: a dominant part scaling with $a_L^0$, that gives the linearized Einstein equations, and a subdominant part scaling with $a_L^4$.

$(iii)$ The third case does apply to theories where we impose a certain set of constraints $z$ weakly. For the linearized theory we have a bare amplitude $\exp(\imath S_0)$ as well as a Gaussian constraint term $\exp(- G)$. $S^0$ is a quadratic functional of $y$ and $z$, and  $G$ is a (positive definite) functional of the $z$ variables only.  We can summarize the $S_0$ bare action and the $G$--part into a complex action $S=S^0 + \imath  G$, and then apply the decoupling transformation (\ref{dtrafo}) for the constraints $z$.  We then obtain an effective complex action for the unconstrained variables $y$
\ba
S'=  \frac{1}{2}\left(  y^\dagger S^0_{yy} y  -  y^\dagger S^0_{yz}  (S^0_{zz}+ \imath G_{zz})^{-1}S^0_{zy}y        \right)
\ea
with complex critical points \cite{EffSF2}
\ba
\left(S^0_{yy}- S^0_{yz}  (S^0_{zz}+ \imath G_{zz})^{-1}S^0_{zy}\right)y=0  \q .
\ea
 The equations of motion for the strongly constrained theory are given by $S^0_{yy}=0$.
Having a scaling $S^0_{yz}$ with a positive power in the lattice constant (relative to the scaling in $S^{0}_{yy}$) does help to keep the corrections to these equations of small. The imaginary parts are also suppressed if $S^0_{zz}$ diverges faster than $G_{zz}$ in the continuum limit.

\section{The linearized Area Regge action on the (tilted) hyper-cubical lattice}\label{Sec:HC}

\subsection{The (tilted) hyper-cubical lattice}

For the expansion of the Area Regge action we will consider a background geometry given by flat Euclidean 4D space. The triangulation will be based on a regular hyper-cuboidal  lattice  with lattice vectors
\ba
 e_0= \lambda(0,0,0,1)+\lambda t \; ,\q e_1=\lambda (0,0,1,0)+\lambda t\; ,\q e_2=\lambda (0,1,0,0)+\lambda t\;,\q e_3=\lambda (1,0,0,0)+\lambda t \q .
 \ea
The vector $t=(s,s,s,s)$ defines a tilting of the lattice.  This tilting allows us to define the linearized Area Regge action. Ultimately, we will be interested in the limit $s\rightarrow 0$.  We apply a periodic identification, so that we obtain a 4-torus with $N^4$ hyper-cubes.
 
To label the vertices of the lattice, we will be using binary notion as introduced in \cite{RocekWilliams}: ignoring the tilting parameter $s$, the labels of the vertices of the hyper-cuboids spanned by the lattice vectors arise by interpreting the vertex coordinates as binary numbers. That is $v_0\equiv \lambda(0,0,0,0)$, $v_1=\lambda(0,0,0,1), v_2=\lambda(0,0,1,0),  v_3=\lambda(0,0,1,1)$ and so on. In the following, we will often refer to vertices by just their labels $v_0=\{0\},v_1=\{1\}$, and to simplices by their vertex sets, e.g. $\{0,7\}$ describes the edge between $\{0\}$ and $\{7\}$. We will always order the labels for a given simplex from smallest to largest.  Hyper-cuboid at vertex $\{X\}$ refers to the hyper-cuboid whose vertex with the smallest numerical value is given by $\{X\}$.

 
 The hyper-cuboid at $\{0\}$ is divided into four-simplices of the form $\{0,2^i,2^i+2^j,2^i+2^j+2^k,15\}$, where $i,j,k$ are pairwise different and can take values $0,1,2$ or $3$. This gives $24=4\cdot3\cdot2$ four-simplices.  All four-simplices are isometric, even on the tilted lattice. This subdivision of the hyper-cube into four-simplices  does also define the number of edges and triangles in the hyper-cube, namely 65 and 110 respectively. 
 
Each simplex of this triangulation is associated to a specific vertex: a simplex with vertex labels $\{X,Y,\ldots \}$ is associated to the vertex $\{X\}$. With this convention we have 24 four-simplices,  60 tetrahedra, 50 triangles and 15 edges associated to a given vertex. Hence we have 50 area variables and 15 length variables per vertex. Thus we should expect 35 area-length constraints per vertex. In the limit $s\rightarrow 0$ the vector describing the hyper-diagonal length in terms of areas is going to zero. This leads to 14 length variables and 36 constraints.

\subsection{Linearized Area and Length Regge calculus on the (tilted) hyper-cubical lattice}

\subsubsection{The Area Regge Hessians for the simplices}

We will next discuss the computation of the Area Regge and Length Regge Hessians on the tilted hyper-cubical lattice.

We will use as variables the squared areas. The fluctuation variable  $\alpha_t$ associated to a triangle $t$ is therefore defined by $A^2_t= {\bf A}_t^2 + \alpha_t$ where ${\bf A}_t^2$ is the background value of the squared area for the triangle $t$.  On  a background with vanishing deficit angles  the quadratic part of the action is given by 
\ba\label{SAR2}
 S^{(2)}_{AR}=\tfrac{1}{2\kappa} \sum_{t,t'} \alpha_t  H_{tt'}\alpha_{t'}  \q \text{with} \q H_{tt'} = - \sum_\sigma \sum_{e\in \sigma}\frac{1}{2A_t}  \frac{\partial \theta_t}{\partial L^2_e}    (J^\sigma)^{-1}_{et} \frac{ \partial L^2_e}{\partial A^2_t} \q 
\ea
with $\kappa=8\pi G/c^3$ and where $J^\sigma_{te}=\partial A_t^2/\partial L^2_e$ is the Jacobian, formed from the lengths square derivatives of the area squares, and associated to the simplex $\sigma$. Importantly, the Hessian is given by a sum over the four-simplices. We can thus compute the Hessian separately for each simplex, and then sum the contributions over all four-simplices in the lattice.

\subsubsection{The hyper-cubical constraints}

In (\ref{SAR2}) we defined the derivatives of the dihedral angles with respect to the squared areas via the derivatives of the dihedral angles with respect to the length squares.\footnote{An alternative possibility to compute the area derivatives of the dihedral angles is via the derivatives of the Angle Gram matrix with respect to the dihedral angles. See \cite{ADH1}.}  Explicit formulas for $\partial \theta^\sigma_t/\partial L^2_e$ can be found in \cite{DittrichFreidelSpeziale,EffSF3}. 

To compute $\partial \theta^\sigma_t/\partial A^2_t$ we therefore need the inverse of the area--length Jacobian $J^\sigma_{te} =\partial A^2_t /\partial L_e^2$ for each four-simplex. These Jacobians are invertible for the simplices of the tilted lattice. But for the tilting parameter $s$ going to zero, $J^\sigma_{te}$ has, for each four-simplex $\sigma$, one left null vector $n^L=\{n^L_t\}_t$ and one right null vector $n^R=\{n^R_e\}_e$.  

The right null vectors $n^R_e=\delta_{e,\{0,15\}} $ describe the fluctuation of the hyper-diagonal in lengths square variables, correspondingly we can associated the left null vectors to the fluctuation of the hyper-diagonal in area square variables. All of the 24 four-simplices in a  given hyper-cube contain the hyper-diagonal, so we have  24 left and 24  right null vectors.  The 24 left null vectors are however all the same --- namely given by the vector describing the length fluctuation of the hyper-diagonal. In contrast, the 24 right null vectors define an independent set if one just considers one hyper-cube.  On the entire lattice one has one dependency per vertex between the entire set of null vectors, that is there are 23 independent right null vectors per vertex.

Despite these null vectors for the area-length Jacobians, one can define linearized Area Regge calculus also on the (not-tilted) hyper-cubical lattice \cite{ADH1}.  The key is to project out all the area fluctuations that correspond to the null vectors $n^L_t$.  The area--length Jacobians are then invertible on the resulting subspace. The null vectors can thus be understood to define constraints for the area fluctuations. We will refer to these constraints as hyper-cubical constraints.

As mentioned above the fluctuations described by these hyper-cubical constraints  correspond to computing the hyper--diagonal length from the different 4-simplices. The area--length constraints ensure that all the different ways of computing a length from the areas coincide. The hyper-cubical constraints are therefore almost all contained in the area--length constraints. There could be one exception --- namely the degree of freedom corresponding to the hyper-diagonal fluctuation. The  derivatives of all the areas with respect to the hyper-diagonal  are however vanishing in the $s\rightarrow 0$ limit, that is in this limit the hyper-cubic constraints do not contain any length degree of freedom.
(For the same reason all the matrix elements associated to the hyper--diagonal  do  vanish in the Length Regge Hessian.)

The hyper-cubical constraints are also significant for the tilted lattice: doing a series expansion in $s$, the inverse Jacobian will include $s^{-1}$ terms. These are given by 
\ba
(J^\sigma)^{-1}_{et} =   \frac{1}{s} \, \delta_{e,\{0,15\}} \, n^L_t +{\cal O}(s^0)         \q .
\ea
As the derivatives $\partial \theta_t^\sigma/\partial L_{\{0,15\}}^2$ and the inverse area squares are of order ${\cal O}(s^0)$, these $s^{-1}$ terms in the inverse Jacobians will descend to $s^{-1}$ terms for the Area Regge Hessians.  Taking the $s\rightarrow 0$ limit, these terms would diverge. Thus  the equations of motion would suppress the degrees of freedom described by the null vectors $n^L_t$.

To take this suppression into account, one can construct an orthogonal projector  $P_{\rm HC}$ (see Appendix \ref{App:Basis}), which projects these null vectors out. Note that it is not sufficient to just subtract the $s^{-1}$ terms, as the Hessian does also include terms of order $s^0$, which are proportional to the constraints, and which also need to be projected out. 

\subsubsection{The Area Regge Hessian on the lattice}

Proceeding with the tilted lattice we can compute the Area Regge Hessian for all 24 simplices of the hyper-cube, and sum these up to get the Hessian for the hyper-cube. To express the Hessian on the entire lattice we employ a lattice Fourier transform: 
\ba\label{FTrafo}
\alpha_t(k')=\sum_\nu \exp( -\imath\,  \Lambda\, \sum_i k'_i\cdot \nu_i ) \alpha_t(\nu) \q ,
\ea
where we introduced a shifted momentum variables $k'=(k'_3,k'_2,k'_1,k'_0)$ with $k'_i=2\pi \kappa'_i/(N \lambda)$ and $\kappa'_i=0,1,\ldots,N-1$ for $i=0,1,2,3$.
 The shifted momenta $k'_j$ are related to the physical momenta $k_j$ by $k'_j=k_j+s  \sum_{i=0}^3 k_i$. 
 
 These shifted momenta arise due to the tilted lattice, whose vertices are at coordinates $\lambda(\nu'_3,\nu'_2,\nu'_1,\nu'_0)$ with $ \nu'_j= \nu_j+ s \sum_{i=0}^3\nu_i$ where $\nu_i=0,1,\ldots,N-1$. We have then $\nu' \cdot k =\nu \cdot k'$.  The Fourier transformation (\ref{FTrafo}) takes into account the   periodicity of the area variables, which for the $\nu$--labels is defined by identifying $\nu_i\equiv  \nu_i + N$. This is achieved by appropriately shifting  the phase factors, so that these have the same periodicity \cite{Hol3D4D}.
 
 Note that the shifted momentum is equal to the physical momentum up to an ${\cal O}(s^1)$ term. For the terms of lowest order in $s$ we can therefore identify the shifted with the physical momentum.

The index $t$  in $\alpha_t(k')$ runs through all 50 triangles associated to a given vertex. We will set $\Lambda$ eventually equal to the background lattice constant $\lambda$. We will keep $\Lambda$ for now, as it allows us to count the powers of the momenta $k$ in a given expression.  The continuum limit is defined as $\lambda\rightarrow 0, N\rightarrow \infty$ so that $N   \lambda$ remains constant.

We therefore have two different length scales. First the lattice constants $\lambda$ and $\Lambda$, where we use the powers in $\Lambda$ to count the number of derivatives. Secondly the tilting parameter $s$ of the lattice. It will be essential, how the matrix elements of the Hessians scale in $\Lambda$ and $s$. The scaling in $\lambda$ will be (mostly) homogeneous, and can of course also be changed by rescaling the variables. 

To be clear, if we say that a certain block of the Hessian scales with $\Lambda^a$  or $s^b$ it means that all matrix elements in this block are at least of order ${\cal O}(\Lambda^a)$ or ${\cal O}(s^b)$ respectively.  Sometimes, if we need to invert a certain block --- which often is only possible perturbatively in either $s$ or $\Lambda$ --- we need to know that all eigenvalues scale in a certain way. In this case we will explicitly say so. 

~\\

The Fourier transformed Hessian can be encoded into a $50\times 50$ momentum--dependent Hermitian matrix.  It has the following properties: 
\begin{itemize}
\item The matrix elements of the Hessian, if not vanishing, include generally terms of order $\Lambda^0$. But many matrix elements also include higher order in $\Lambda$--terms, that is derivatives.  The Hessian includes a global $\lambda^{-6}$ factor.
\item  Some matrix elements of the Hessian include $s^{-1}$ terms. The $s^{-1}$  part of the Hessian defines a strictly positive quadratic form on the space spanned by the hyper-cubic constraints. That is  all eigenvalues of this $s^{-1}$ part are strictly positive. 

\item There is a $8\times8$ block $B^{\rm A}_{\rm SP}=-\lambda^{-6}\frac{4}{(1+4s)^3} \mathbb{I}_8$ for the variables corresponding to 8 of the bulk triangles in the hyper-cube. This block does therefore not contain $s^{-1}$ terms. The hyper-cubic constraints are indeed orthogonal to the vectors describing the area square fluctuations of these 8 bulk triangles. There are, however, couplings between these 8 variables and the remaining 42 variables.

 These 8 bulk triangles are given by two sets of four triangles: $\{\{0,2^i,15\}\}_{i=0}^3$ and $\{\{0,15-2^i,15\}\}_{i=0}^3$. (Notice, that the two triangles $\{0,2^i,15\}$ and $\{0,15-2^i,15\}$ are lying in the same plane spanned by the edges $\{0,2^i\}$ and $\{0,15-2^i\}$, that is $e_i$ and $e_j+e_k+e_l$ with $j,k,l\neq i$.)

The  block $B^{\rm A}_{\rm SP}$ is local, that is, does not include a momentum dependence.  One can thus integrate out the 8 degrees of freedom, without introducing any inverse momenta.  After this step there is another $6\times6$ block, which, although not diagonal (and including $1/s$ terms), does also not include any momentum dependence, and can therefore also be integrated out without introducing inverse momenta. This block is associated to the remaining 6 inner triangles in the hyper-cube. We will however see, that the degrees of freedom associated to the first set of 8 triangles play a special role, and are, indeed, spurious variables.

\item There are four, and only four, null-vectors corresponding to vertex translations, that is lattice diffeomorphisms \cite{RocekWilliams,DiffReview08}. These are given by 
$
n^i_t=\sum_e \frac{\partial A^2}{\partial L^2_e} n^i_e
$
where $n^i_e$ describes a vertex translation in length square variables in the direction of  $\{0,2^i\}$ with $i=0,1,2,3$. 
\end{itemize}

\subsubsection{Geometric interpretation of the Area Regge Hessian and Area Regge degrees of freedom}

Note that the Area Regge Hessian is given by
\ba
H_{tt'} =
 \frac{1}{2A_t}  \frac{\partial \epsilon_t}{\partial A^2_{t'}} 
 \ea
  that is the rows define (rescaled) linearized deficit angles in the area square variables.  As the Hessian has null vectors, which describe the linearized diffeomorphisms, we can conclude that the linearized deficit angles provide a basis of variables, that are invariant under the linearized lattice diffeomorphisms.  There are $50-4=46$ such independent variables per lattice vertex.

We can contract the Area Regge Hessian with the area-length Jacobian $\partial A_t^2/\partial A^2_e$ and now obtain the (rescaled) linearized deficit angles as function of the length square fluctuations
\be
\tilde \varepsilon_t = \frac{1}{2A_t}  \frac{\partial \epsilon_t}{\partial L^2_e}  \q .
\ee
As pointed out by Barrett \cite{BarrettWilliams}, these deficit angles $\tilde \varepsilon_t$ are not independent. Rather there are three Bianchi identities for the linearized deficit angles associated to triangles sharing the same edge.  That is, there are three Bianchi  null vectors $(b^e_\alpha)_t$, with $\alpha=1,2,3$, for each edge. These are left null vectors for the matrix formed by the $\tilde \varepsilon_t$. They  can be easily found by isolating the set of triangles ${\cal S}_t(e)$  which share a given edge, and determining the left null vectors for the matrix formed by the set $\{\tilde \varepsilon_t , t\in {\cal S}_t(e)\}$. For the hyper-cubical lattice the Bianchi identities are also listed in \cite{BarrettWilliams}.

Note however that these Bianchi identities do not hold for the linearized deficit angles in area variables.  We can in fact find a basis of area-length constraints in the following way. We collect all Bianchi  null vectors $(b^e_\alpha)_t$ and form
\be\label{Bconstraints}
(c^e_\alpha)_t=(b^e_\alpha)_t  H_{tt'}   \q .
\ee
By construction, these vectors do vanish if contracted from the right with a vector  $\partial A^2/\partial L_e^2$ describing a length fluctuation. Hence these vectors can be understood as conditions on the area--fluctuations. If these conditions are satisfied the area fluctuations result from a length fluctuation. 

We have 15 edges per vertex, so a priori we get $3\times 15=45$ constraints.  There are however 6 dependencies between the null vectors $(b^e_\alpha)_t$, giving rise to 6 dependencies for the constraints (\ref{Bconstraints}). Additionally $H_{tt'}$ has itself four null vectors, given by the lattice diffeomorphisms. These lead to another 4 dependencies, which are independent from the previous ones. We are left with 35 independent constraints --- and this is indeed the correct number to reduce the 50 area variables to the 15 length variables per vertex.

Why are the Bianchi identities not satisfied for the deficit angles in area variables? We should note that the Bianchi identities for the $\tilde \varepsilon_t$ correspond to the torsion--free Bianchi identities of the continuum. It has been argued in \cite{DittrichRyan,TorsionConnection,BFCG2} that the area--length constraints can be identified with torsion. 
In the presence of torsion the Bianchi identities relate curvature and torsion, which explains why the torsion free version is not satisfied in Area Regge calculus. Matching the constraints (\ref{Bconstraints}) with the torsion modified Bianchi identities could help to identify a lattice version of torsion.

\subsubsection{The Length Regge Hessian on the lattice}\label{Sec:LRegge}

The fluctuation associated to the length square of an edge $e$ is described by the vector $(\ell _e)_t=\partial A^2_t/\partial L^2_e$. Contracting the Area Regge Hessian on both sides with these vectors we obtain the Length Regge Hessian $H^L$. The $(\ell_e)_t$ include the area-length Jacobian, which cancels the inverse area-length Jacobians, which we used in the construction of the Area Regge Hessian. As the inverse Jacobians gave rise to the $1/s$ terms, we will not encounter any such terms in the Length Regge Hessian.  We can in particular set the tilting parameter $s$ to zero. This then results in the Length Regge Hessian for the hyper-cubical lattice, as first computed by Rocek and Williams in \cite{RocekWilliams}. 

Note  that the vectors $\ell_e$, for $s=0$ are orthogonal to the hyper-cubical constraints. If $s=0$ one has only 14 such vectors, as the area fluctuation describing the hyper-diagonal is vanishing in the limit $s\rightarrow 0$. Furthermore, we note that for $s=0$ the vectors describing  body--diagonals $\{0,15-2^i\}_{i=0}^3$, only include area fluctuations, associated to the set of 8  triangles $\{\{0,2^i,15\}\}_{i=0}^3$ and $\{\{0,15-2^i,15\}\}_{i=0}^3$, that is only include spurious variables.

The Length Regge Hessian for the hyper-cubical lattice (with $s=0$) has the following properties:
\begin{itemize}
\item The Length Regge Hessian has a global scaling factor $\lambda^{-2}$. Non-vanishing matrix elements start  with $\Lambda^0$ terms, but can also include higher order terms in $\Lambda$.
\item All entries associated to the hyper-diagonal are vanishing.
\item The block describing the body--diagonals is given by $H^{\rm L}_{\rm SP}=-\frac{1}{2}\lambda^{-2} \,\mathbb{I}_4$. 
\item Apart from the hyper-diagonal, the Hessian has four further null vectors $(n^i)_e$, describing vertex translations, that is lattice diffeomorphisms. 
\end{itemize}

As described in Section \ref{Sec:Decoupling}, we can apply a decoupling transformation for the four variables describing the body--diagonals. The transformed Hessian is then given by two decoupled blocks, the first one $H^L_{M}$ is a $10\times 10$ block, the second one is the block $H^{\rm L}_{\rm SP}$. 

 One will then find that all terms in the $H^L_{M}$ block are at least of order $\Lambda^2/\lambda^2$. Note that this is the case only after the decoupling transformation. The $H^{\rm L}_{\rm SP}$ scales with $\Lambda^{0}/\lambda^2$, that is the body--diagonal variables resulting from the decoupling transformations   have to vanish in the continuum limit\footnote{This statement has to be understood in relation to the variables describing the remaining length fluctuation. The Hessian block for these variables is of order ${\cal O}(\Lambda^2/\lambda^2)$, that is remains finite if we send $\Lambda=\lambda$ to zero.}, if we demand that the action remains finite. 

The body--diagonal variables can be interpreted as spurious, as they are not needed to define the metric fluctuation variables from the length square  fluctuation variables. Indeed the lengths squares of the  edges $\{0,2^i\}_{i=0}^3$ define the diagonal metric fluctuations $\ell^2_{\{0,2^i\}}=h_{ii}$, whereas the lengths squares of the face diagonals $\{0,2^i+2^j\}_{i>j}$ are given by $\ell^2_{\{0,2^i+2^j\}}= h_{ii}+h_{jj}+2h_{ij}$ and thus encode all non-diagonal metric fluctuations.

Transforming the $10\times 10$ block of the Length Regge Hessian  to metric fluctuation variables, we find a discretization of the quadratic expansion of (minus) the Einstein-Hilbert action. More precisely, we can introduce spin zero mode and spin two mode projectors ${}^0\!\Pi_{\rm L},{}^2\!\Pi_{\rm L}$ on the lattice \cite{BahrDittrichHe}.  Setting $\Lambda=\lambda$, the  Length Regge Hessian, transformed to metric variables, is then given by 
\ba\label{h-sec}
H^L_M= -\frac{1}{4}\Delta_{\rm L} ({}^2\!\Pi_{\rm L}-2\,{}^0\!\Pi_{\rm L})
\ea
with 
\ba
\Delta_{\rm L}=\frac{1}{\lambda^2}\sum_{i=0}^3 \left(2-\exp(\imath \Lambda k_i^2)-\exp(-\imath \Lambda k_i^2)\right)\,\,=\, \, \frac{\Lambda^2}{\lambda^2}\sum_{i=0}^3 k_i^2 + \frac{1}{\lambda^2} {\cal O}(\Lambda^3)
\ea
 the lattice Laplacian. (Here we get minus the linearized Einstein-Hilbert action. This is due to the standard definition of the Regge action in (\ref{Eq:LR}), which differs by a global sign from the Euclidean gravity action.)

\subsubsection{Separation of variables}\label{Sec:separation}

We have already defined two sets of fluctuation variables: Per lattice vertex we have  $(a)$ the 14 variables corresponding to the length fluctuations of all edges associated to this vertex, except the hyper-diagonal, and $(c)$ 
the 23 variables given by the hyper-cubical constraints. We thus need to still identify a set $(b)$ of 13 variables, which describe a set of area-length constraints in addition to the hyper-cubical constraints. We will choose these constraints to be orthogonal to the hyper-cubical constraints in the $s\rightarrow 0$ limit and refer to these variables as $\zeta$--variables.

To identify the $\zeta$--variables it is helpful to first `remove' the hyper-cubical constraints. To this end we construct a orthonormal projector $P^{\rm HCC}$ onto the hyper-cubical constraints, see Appendix \ref{App:Basis}. With the exception of the entries for the 8 spurious area variables, all elements of this projector are non-vanishing, if we express it in the $\alpha_t$--variables. Using this projector does quickly lead to very involved expressions. 

A more convenient procedure is to find a basis, which makes the split of the variables into the sets $(a)$, $(b)$ and $(c)$ explicit. The projector $P^{\rm HCC}$ is however quite helpful, to identify such a basis. 

The construction of this basis involves the following steps:
\begin{itemize}
\item First, we consider the set of 50 triangles $\{0,X,Y\}$ associated to the vertex $\{0\}$. Remember that we only consider ordered tuples of vertex indices.  $X$ and $Y$ are of the form $\sum_{i=0}^3 2^i b_i $ with $b_i$ either being $0$ or $1$. If $b_i$ is $1$ for $X$ it also has to be $1$ for $Y$ and $Y>X$. 

We notice that the triangles associated to a given vertex form pairs, where the triangles of a given pair are in the same plane. E.g. $\{0,1,3\}$ and $\{0,2,3\}$ are in the plane spanned by $\{0,1\}$ and $\{0,2\}$. Generally the pairs are of the form $\{0,X,X+X'\}$ and $\{0,X',X+X'\}$.   This suggest to introduce new variables 
\ba\label{trafo1}
\alpha^+_{(X,X')}=\frac{1}{2} \left(\alpha_{\{0,X,X+X'\}}+ \alpha_{\{0,X',X+X'\}} \right) \, ,\q 
\alpha^-_{(X,X')}=\frac{1}{2} \left(  \alpha_{\{0,X,X+X'\}} - \alpha_{\{0,X',X+X'\}} \right) \q.
 \ea
 defined for the following pairs $(X,X')$ (indices $i,j,k,l$ run all from $0$ to $3$):
 \begin{itemize}
 \item 6 planes $(X,X')$ with $X=2^i$ and $X'=2^j$, with $j>i$. These planes are spanned by pairs of principal axis vectors $(e_i,e_j)$. 
 \item 12 planes $(X,X')$ with $X=2^i$ and $Y=2^j+2^k$ where $k,j\neq i$ and $k>j$. These planes are spanned by a principal axis vector $e_i$ and a  face diagonal $e_j+e_k$ (orthogonal to $e_i$ in the $s\rightarrow 0$ limit).
 \item 3 planes $(X,X')$ with $X=2^0+2^i$ and $X'=2^j+2^k$ with $i,j,k>0$ and pairwise different. These planes are spanned by pairs of face diagonals $(e_0+e_i, e_j+e_k)$ (orthogonal to each other in the $s\rightarrow 0$ limit).
 \item 4 planes $(X,X')$ with $X=2^i$ and $X'=2^j+2^k+2^l$ with $j,k,l \neq i$ and $j>k>l$. These planes are spanned by a principal axis vector $e_i$ and a body diagonal $e_j+e_k+e_l$ (orthogonal to $e_i$ in the $s\rightarrow 0$ limit). The $(4+4)$ area variables associated to these planes are spurious variables.
 \end{itemize}
 See also Appendix \ref{App:Basis} for an explicit listing of the 25 pairs $(X,X')$.

 This transformation does indeed decouple the $\alpha^+$ variables from the $\alpha^-$ variables in the projector $P_{\rm HCC}$.  The projector can be even made free of any $k$--dependence by adjusting slightly the definition of the $\alpha^+$ and $\alpha^-$ variables. In (\ref{trafo1}) we consider pairs of triangles which are in the same plane and associated to the same vertex. We could also consider triangles associated to different vertices -- this would introduce (in the Fourier transform) phase-factors in the transformation. The following choice does make $P_{\rm HCC}$ free of any momentum dependence:
\ba\label{trafo2}
\alpha^+_{(XX')}=\frac{1}{2} \left( \omega_X \,\alpha_{\{0,X,X+X'\}}+  \omega_{X'}  \,\alpha_{\{0,X',X+X'\}} \right)  \, ,\q 
\alpha^-_{(XX')}=\frac{1}{2} \left(  \omega_X \, \alpha_{\{0,X,X+X'\}} -   \omega_{X'} \, \alpha_{\{0,X,X+X'\}} \right)  ,\;\;\;
\ea
where $\omega_{X}$ with binary label $X=\sum_i b_i 2^i$ is given by $\omega_{X}= \exp(\imath \Lambda \sum_i b_i k'_i )$. 
\item We separate with this transformation the variables into an $\alpha^+$  and an $\alpha^-$ set, which contains each $25=21+4$ variables, where we singled out  $(4+4)$ spurious fluctuations $\alpha^{\pm}_{(2^i,2^j+2^k+2^l)}$. In fact we can now explain why one can interpret these variables as spurious: the $\alpha^+_{(X,X')}$ variables give as the area (squares) fluctuations of parallelograms  spanned by edges $\{0,X\}$ and $\{0,X'\}$. This information can be also encoded into an area metric \cite{Schuller}, which has\footnote{These 21 components arise from the following counting. Consider double indices $I=(ij)$, with $i<j$ and $i,j=0,\ldots,3$. There are six such double indices. The area metric tensor $G_{IJ}$ is symmetric under exchange of $I$ and $J$. This gives $(6\times 5/2)+6=21$ components.} 21 components.  Similar to the relation between the length variables and the (length) metric, we only need 21 of the areas to define the area metric components. 

We will however see that the hyper-cubic constraints do suppress all fluctuations in the area metric variables that are not coming from length metric fluctuations. 

\item
We mentioned above that the projector onto the hyper-cubic constraints does separate into an $\alpha^+$ block and an $\alpha^-$ block. Additionally, it does not involve the spurious variables. The 23 hyper-cubic constraints split into 11 conditions for the 21 (non-spurious) $\alpha^+$  variables and 12 conditions for the 21 (non-spurious) $\alpha^-$  variables. Appendix \ref{App:Basis}  lists a basis for these constraints.

\item
As a next step we should identify the set $(a)$ and $(c)$ of variables, that is the length variables and the $\zeta$--variables.  
To this end remember that the length variables are orthogonal to the hypercubic constraints in the $s\rightarrow 0$ limit. The split of the areas into $\alpha^+$ and $\alpha^-$ parts can be used to also split the lengths variables: $\ell_e=\ell^+_e(\alpha^+)+ \ell^-_e(\alpha^-)$.

Importantly, as the hypercubic constraints do not couple the $\alpha^+$ and the $\alpha^-$ variables, all 14 $\ell^+_e$ and 14 $\ell^-_e$ vectors are orthogonal to the hypercubic constraints in the $s\rightarrow 0$ limit. The $\ell_e^+$ vectors turn out to be linearly independent, whereas there is one dependency (in the $s\rightarrow 0$ limit) between the 14 $\ell^-_e$ vectors.  
Below we will choose an independent  basis of 13 vectors $\{\zeta_a\}_{a=1}^{13}$, which will span the space of  $\ell^-_e$ vectors.

\item 
We noted that the $\ell_e^-$ vectors are not linearly independent. Additionally each $\ell_e^-$ comes with a $\Lambda^1$-- scaling, which results from the $\omega$--shifts in the definition (\ref{trafo2}) of the $\alpha^+$ and $\alpha^-$ variables. But it turns out, that in the $s\rightarrow 0$ limit,  the $\omega$--dependent factor is the same for each component $(\ell_e^-)_t$ of a given edge, so we can just remove these factors.\footnote{For a calculation to $s^1$ order one might not want to remove these $\omega$ factors, as it would lead to ${\cal O}(s^1)$ terms involving quotients of momenta.} We keep the $\lambda^2$ scaling, which results from the area square -- length square Jacobian. In this way the $\ell_e$ vectors and the $\zeta$--vectors will scale in the same way. We give an explicit choice of basis for the $\zeta$--sector in Appendix \ref{App:Basis}. This choice does lead to  a very elegant expression for the Hessian blocks involving the $\zeta$-variables. This basis does also include four vectors which include only spurious degrees of freedom.

\end{itemize}

In summary we split the 50 area square fluctuations per vertex into the following sets of variables:
\begin{itemize}
\item[$(e)$]
 A set of 14 vectors $\ell_e$ describing the squared edge lengths. 10 of these correspond to the length metric fluctuations, and 4 are spurious variables.

\item[$(b)$]
A set of 13  vectors $\zeta_b$, which only include $\alpha^-$ variables. 4 of these are spurious variables.
\item[$(c)$]
A set of 23 hyper-cubical constraints $\{\chi_d\}_{d=1}^{23}$. These split into 11 constraints for the $\alpha^+$ sector and 12 constraints for the $\alpha^-$ sector.
\end{itemize}

We perform a further transformation for the sets $(a)$ and $(b)$:  We compute the Area Regge Hessians  restricted to the $(a)$ variables, that is 
\ba
H_{ee'}=\sum_{tt'} (\ell_e^\dagger)_t H_{tt'} (\ell_{e'})_{t'}  = H^{\rm LR}_{ee'}
\ea
 which coincides with the Length Regge Hessian. Likewise we compute the Area Regge Hessian restricted to the $(b)$ variables 
 \ba
 H_{bb'} = \sum_{tt'}  (\zeta_b)_t H_{tt'} (\zeta_{b'})_{t'}  \q .
\ea
 Separately, for both sets we apply a  transformation that decouples  the spurious variables from the remaining ones. These transformations do only redefine the spurious variables.

Thus we have now the following sets of variables
\begin{itemize}
\item[$(a_1)$] A set of 10 vectors $\ell_e=\ell_e^+(\alpha^+)+\ell_e^-(\alpha^-)$, where $e$ runs through the four principal axis vectors and the six face diagonals. We can transform these vectors such that they describe fluctuations in the 10 independent metric components, see Section \ref{Sec:LRegge}. 
\item[$(a_2)$] A set of 4 vectors $\ell'_e$, associated to the body diagonals, describing spurious variables.  
\item[$(b_1)$] A set of 9  vectors $\zeta_{b_1}$, $b_1=1,\ldots,9$, only involving $\alpha^-$ variables.
\item[$(b_2)$] A set of 4 vectors  $\zeta'_{b_2}$, $b_2=1,\ldots 4$, only involving $\alpha^-$ variables and describing spurious variables. 
\item[$(c_+)$] A set of 11 hyper-cubical constraints $\{\chi_{c_+}\}_{c_+=1}^{11}$ in the $\alpha^+$ variables. 
\item[$(c_-)$] A set of 12 hyper-cubical constraints $\{\chi_{c_-}\}_{c_-=1}^{12}$ in the $\alpha^-$ variables. 
\end{itemize}

Vectors  from the set $(c)=(c_+) \cup (c_-)  $ are, in the $s\rightarrow 0$ limit, orthogonal to the vectors in $(a)$ and $(b)$. But vectors from $(a)$ are not necessarily orthogonal to vectors from $(b)$.

In the following we denote with $H_{({\cal S}_1) ({\cal S}_2)}$ the Area Regge Hessian contracted from the left with the adjoints of the vectors in the set ${\cal S}_1$ and contracted from the right with the vectors in the set ${\cal S}_2$. We will also use abbreviations for the union of sets, e.g. $(a)=(a_1)\cup (a_2)$,$(b)=(b_1)\cup (b_2)$ , $(ab)=(a)\cup (b)$ and $(c)=(c_+)\cup (c_-)$

\subsubsection{Decoupling the hyper-cubical constraints}

Next we will apply decoupling transformations for the variables describing the area--length constraints. We begin with the hyper-cubical constraint set $(c)=(c_+)\cup (c_-)$. 

All 23 eigenvalues of the ${\cal O}(\Lambda^0)$ order of the Hessian block $H_{(c)(c)}$ are non-vanishing and scale homogeneously with $s^{-1}$. We can therefore invert $H_{(c)(c)}$ perturbatively in $\Lambda$, which will result in an inverse of order ${\cal O}(s^1)$. On the other hand, we have that the Hessian blocks $H_{(a)(c)}$ and $H_{(b)(c)}$ do not involve $s^{-1}$ terms, that is are at least of order $s^0$.\footnote{The $(a)$ and $(b)$ vectors are (for $s=0$) orthogonal to the hyper-cubical constraints. The columns (or rows) of the $s^{-1}$ part of the Hessian give the hyper-cubical constraints (or adjoints) thereof. Thus contracting the Hessian from either side with vectors from $(a)$ or $(b)$ projects out all $s^{-1}$ terms from the Area Regge Hessian.}

Applying the decoupling transformation described in Section \ref{Sec:Decoupling}  for the hyper-cubical constraints, we will find  a modifiction for the (effective) action for the remaining variables, given by
\ba
H'_{(ab)(ab)}= H_{(ab)(ab)}- H_{(ab)(c)} H^{-1}_{(c)(c)} H_{(c)(ab)}  \q .
\ea
The scaling behaviour in $s$ of the various blocks, described above, ensures that the correction $(H'_{(ab)(ab)}-H_{(ab)(ab)})$ is of order ${\cal O}(s^1)$. Hence this modification vanishes in the $s\rightarrow 0$ limit. The block $H_{(ab)(ab)}$ is, on the other hand, of order ${\cal O}(s^0)$. Next we will decouple the $\zeta$--variables, which requires to invert the $H_{(b)(b)}$ block. The eigenvalues of the $s^0 \Lambda^0$ part of this block are all non-vanishing. For the $s\rightarrow 0$ limit we therefore need to only consider the $s^0$--parts in $H_{(ab)(ab)}$. That is we can finally set $s=0$.

\subsubsection{Decoupling the $\zeta$--variables}\label{Sec:zeta}

We can now consider the decoupling transformation for the $\zeta$--variables. We can perform this transformation perturbatively in an expansion in $\Lambda$, that is in derivatives. The Hessian blocks in $(a_1),(a_2),(b_1),(b_2)$ are of the following $\Lambda$--order:
\ba\label{Scaling1}
H_{(ab)(ab)}\,\sim \q \frac{1}{\lambda^2}\,\,\times\,\,
\begin{tabular}{|c |c|c|c|c|}
\hline
&$(a_1)$ & $(a_2)$ &$ (b_1)$ & $(b_2)$ \\   \hline
$(a_1)$ & $\;\Lambda^2\;$ & $ 0$ &  $\;\Lambda^3\;$ & $0 $ \\ \hline
$(a_2)$ & $ 0$ & $\Lambda^0$ & $\Lambda^1$ & $\;\Lambda^1\;$\\ \hline
$(b_1)$ & $\Lambda^3$& $\Lambda^1$ & $ \Lambda^0$ & $0 $\\\hline
$(b_2)$ & $0$ & $\;\Lambda^1\;$ & $0$ & $ \Lambda^0$ \\ \hline
\end{tabular}
\ea

In particular all eigenvalues of $H_{(b)(b)}$    start with $\Lambda^0/\lambda^2$ terms and are all strictly negative at this order.  

The three surprising features in (\ref{Scaling1}) are that $(i)$ the $H_{(a_1)( b_1)}$ block is of order ${\cal O}(\Lambda^3)$ and $(ii)$ that the $H_{(a_1)( b_2)}$ block is vanishing to all orders in $\Lambda$ and $(iii)$ that the $H_{(b_1)(b_1)}$ block is of order ${\cal O}(\Lambda^0)$ (and thus not of the same order as $H_{(a)(a)}$).

With this information on the scaling in $\Lambda$ we can determine the (minimal) $\Lambda$--order for the $\zeta$--induced correction for the length sector. This correction is given by
\ba
\Gamma_{(a)(a)} = - H_{(a) (b)} \cdot  H_{(b)(b)}^{-1} \cdot H_{(b)(b)} \q .
\ea
With the scalings in (\ref{Scaling1}) we therefore expect that 
\ba\label{Scaling2}
\Gamma_{(a)(a)} \sim \q \frac{1}{\lambda^2}\,\,\times \,\,
\begin{tabular}{|c |c|c|} \hline
&$(a_1)$          & $(a_2)$ \\\hline
$(a_1)$ & $\Lambda^6$  & $ \Lambda^4  $ \\\hline
$(a_2)$ & $\Lambda^4$  &$ \Lambda^2 $ \\\hline
\end{tabular}         \q .
\ea
We indeed find this scaling behaviour for $\Gamma_{(a)(a)}$. (That is there are no cancellations of terms leading to a scaling in higher order than the minimal possible ones given in (\ref{Scaling2}).) The modified Hessian for the length degrees of freedom is then given by $H'_{(a)(a)}=H_{(a)(a)}+ \Gamma_{(a)(a)}$.  Whereas the spurious degrees of freedom decouple in $H_{(a)(a)}$, that is $H_{(a_1)(a_2)}=0$, this is not the case for $\Gamma_{(a_1)(a_2)}$. We could therefore apply a further decoupling transformation for the $(a_2)$ variables. But as $H_{(a_2)(a_2)}$ is of order $\Lambda^0$, the lowest order modification for the $((a_1)( a_1))$--block is of order $\Lambda^8$. To obtain the lowest order  $\Lambda^6$ correction for $\Gamma_{(a_1)(a_1)}$, we therefore do not need to perform this additional decoupling transformation.

~\\
Let us emphasize an essential point, namely that the Area Regge Hessian itself does already separate the length degrees of freedom from the $\zeta$--degrees of freedom, and within the set of length degrees of freedom, the 10 metric degrees of freedom, from the 4 spurious variables.

\section{The $\zeta$--induced modification to the graviton action}\label{Sec:Corr}

Finally, we can compute the modification to the gravitational action. We did establish in the previous section, that the lowest order for this modification is of order $\Lambda^6/\lambda^2$, and that its computation only involves the blocks of $H$ with respect to the 10 metric variables $(a_1)$ and  the nine $\zeta$--variables $(b_1)$.

This correction is given by 
\ba
\Gamma_{(a_1)(a_1)} =- H_{(a_1)(b_1)}  H_{(b_1)(b_1)} ^{-1} H_{(b_1)(a_1)}    \q .
\ea

The matrix  $H_{(b_1)(a_1)}$ is a $9\times 10$ matrix, which inherits the four right null vectors, describing lattice diffeomorphisms,  from the full Area Regge Hessian. Thus there are also (at least) three left null vectors. Indeed, with our choice of basis (given in Appendix \ref{App:Basis}), the following pairs of rows in $H_{(b_1)(a_1)}$ are equal: (1,2), (3,4) and (5,6).  We can thus introduce a reduced $6\times10$ dimensional matrix $H_{(b'_1)(a_1)}$. We  name the six labels associated to $(b'_1)$ with index pairs   $\{ij\}=\{01\},\{02\},\{12\},\{03\},\{13\}$. We also name the 10 labels of the $(a_1)$ set (after transformation to metric variables with index pairs $\{kl\} = \{00\},\{11\},\{22\},\{33\},\{01\},\{02\},\{12\},\{03\},\{13\}$.

It turns out that (with our choice of basis)  this reduced matrix $H_{(b'_1)(a_1)}$ is given by
\ba
(H_{(b'_1)(a_1)})_{\{ij\} \{kl\}} &=&   4 \imath \Lambda \,(k_i+k_j)\, (H_{(a_1)(a_1)})_{\{ij\} \{kl\}} + 
\frac{1}{\lambda^2} {\cal O}(\Lambda^4) +{\cal O}(s^1)
\ea
where 
\ba\label{EHL}
H_{(a_1)(a_1)} =-\frac{1}{4}\frac{\Lambda^2}{\lambda^2} \Delta ({}^2\!\Pi - 2\,{}^0\!\Pi) +\frac{1}{\lambda^2}{\cal O}(\Lambda^3)
\ea
with $\Delta=\sum_i k_i^2$, is the lattice discretization of the linearized Einstein-Hilbert action (\ref{h-sec}).  That is the $\{ij\}$-row of $(H_{(b'_1)(a_1)})$ is proportional to $(k_i+k_j)$ times the $\{ij\}$-row of  the Hessian matrix for the Einstein-Hilbert action $H_{(a_1)(a_1)}$, for indices $j>i$.

Before proceeding let us comment on what one could have guessed beforehand about the structure of $H_{(b'_1)(a_1)}$. It does inherit the linearized diffeomorphism invariance, that is, its rows should be expressible in a basis of diffeomorphism invariant quantities. The rows of  $(H_{(a_1)(a_1)})$, which give combinations of linearized deficit angles in terms of the metric fluctuations, do provide an over-complete basis of such quantities. The over-completeness results from the linearized diffeomorphism invariance: only 6 of the 10 rows of $(H_{(a_1)(a_1)})$ are independent. The six rows can be chosen to be the ones corresponding to the non-diagonal metric elements $h_{ij}$ with $i<j$.  
We chose the basis for the $\zeta$--vectors such that this match between the rows of $H_{(b'_1)(a_1)}$ and the rows of $H_{(a_1)(a_1)}$ becomes evident. But this choice also happens to maximally simplify  the $\Lambda^0$--part of $H^{-1}_{(b'_1)(b'_1)}$. 

This $\Lambda^0$ part of the reduced version of $H^{-1}_{(b_1)(b_1)}$ is proportional to the identity matrix 
\ba\label{Hbs}
H^{-1}_{(b'_1)(b'_1)} \,=\,  -\frac{\lambda^2}{256} \mathbb{I}_{6}
\ea

This allows us to write the correction as
\ba
(\Gamma_{(a_1)(a_1)})_{\{kl\}\{k'l'\}} = \frac{\Lambda^2\lambda^2}{64}  \sum_{j>i}   (H_{(a_1) (a_1)} )_{\{kl\}\{ij\}} \, (k_i+k_j)^2 \, (H_{(a_1)(a_1)})_{\{ij\}\{k'l'\}} + \frac{1}{\lambda^2}{\cal O}\left(\Lambda^7\right)  +    {\cal O}(s^1)  \q .
\ea

Using (\ref{EHL}) to express $(H_{(a_1) (a_1)} )$ we therefore have
\ba\label{FF}
\Gamma&:=& \sum_{\{kl\},\{k'l'\}} \bar{h}_{\{kl\}}\,(\Gamma_{(a_1)(a_1)})_{\{kl\}\{k'l'\}}  \,h_{\{k'l'\}}  \nn\\
&=&\frac{\Lambda^6}{\lambda^2}\frac{1}{(64)^2}  \, \bar{h}_{kl}  \, ({}^2\!\Pi - 2\,{}^0\!\Pi)^{klij}  \,\, {M_{iji'j'}}^{mn} \, \Delta^2\,k_mk_n \,  \,\, ({}^2\!\Pi - 2\,{}^0\!\Pi)^{i'j'k'l'}\,\, h_{k'l'} \;+ \frac{1}{\lambda^2}{\cal O}\left(\Lambda^7\right)  +    {\cal O}(s^1) \nn\\
&=& \frac{\Lambda^6}{\lambda^2}\frac{1}{(64)^2} \,\;\; \bar{ {\cal G}}^{ij} \,\, {M_{iji'j'}}^{mn} \,\, k_mk_n\,\,  {\cal G}^{i'j'}\;+ \frac{1}{\lambda^2}{\cal O}\left(\Lambda^7\right)  +    {\cal O}(s^1)
\ea
where in the second line we use the Einstein summation convention and treat the indices $i,j,k,l,m,n=0,1,2,3$ as independent. 
In the third line we introduced the linearized Einstein tensor $ {\cal G}^{ij}=\Delta ({}^2\!\Pi - 2\,{}^0\!\Pi)^{ijkl}\,\, h_{kl}$.
The tensor  ${M_{iji'j'}}^{mn}$ has only 1 or 0 as entries: It is equal to 1 if 
\ba
(i\neq j  \land  i'\neq j') \,\, \land \,\,  
\left((i=i' \land j=j')\lor (i=j' \land j=i')\right) \,\, \land \,\, 
\left( (m=i \lor m=j) \land  (n=i \lor n=j) \right) \q ,
\ea
and vanishing in all other cases. We can therefore define ${M_{iji'j'}}^{mn}$  as  (where we are not using Einstein's summation convention)
\ba
{M_{iji'j'}}^{mn}=(\delta_{ii'} \delta_{jj'} +\delta_{ij'} \delta_{ji'})(1-\delta_{ij})(1-\delta_{i'j'}) (\delta_i^m+\delta_j^m)(\delta_i^n+\delta_j^n) \q .
\ea
This is however not a tensor, which is covariant under coordinate transformations. In fact, although the correction (\ref{FF}) is invariant under infinitesimal diffeomorphisms, that is it vanishes for the longitudinal modes,  it does not have the right structure, in order to result from a curvature invariant built from contracting the Riemann tensor, derivatives and the metric tensor.\footnote{I thank Benjamin Knorr for clarifying this point.} It might however come from an invariant involving the Levi-Civita symbol.

In summary, we obtain a correction that is $(i)$ quadratic in the curvature and $(ii)$ of sixth order in the momenta and $(iii)$ after setting $\Lambda=\lambda$ of fourth order in the lattice constant.

Of course the lattice graviton action, which is encoded in $H_{(a_1) (a_1)}$,  has apart from the lowest order term $\Lambda^2/\lambda^2$ also terms of higher order $\Lambda^p/\lambda^2$ with $p=3,4,\ldots$.   These result from an expansion of the difference operators in (\ref{h-sec}) to higher powers in the lattice constant and the momenta $k_i$. 

The $\zeta$--induced correction $\Gamma_{(a_1)(a_1)}$ to (\ref{h-sec}) is therefore of higher order in the lattice constant as compared to the corrections, which result from an expansion of the Length Regge action to higher than second order in $\Lambda$.

\section{Outlook: Weak imposition of area-length constraints}\label{Outlook}

The effective spin foam models \cite{EffSF1,EffSF2,EffSF3} are based on the Area Regge action, but do impose weakly the area--length constraints. The weak imposition of the constraints can be approximated by including a  Gaussian factor, which suppresses fluctuations away from the constraint hypersurface, into the path integral \cite{EffSF1}. This Gaussian factor can be absorbed into the action, if we allow this action to be complex.  The linearized dynamics for this system is then described by a Hessian
\ba
H_C=H+\frac{\imath}{\gamma} G
\ea
where $H$ is the Area Regge Hessian and $G$ a strictly positive bilinear form in the (linearized) area-length constraints.  $\gamma$ is the Barbero--Immirzi parameter, which appears as an anomaly parameter for the area-length constraints. It therefore parametrizes the variance of the Gaussians suppressing the constraint fluctuations \cite{EffSF1,EffSF3}. 

Using the Gaussian factors defined in \cite{EffSF3} we have determined the scaling behaviour of the blocks in $G$ with respect to the splitting of the area variables into the blocks $(a),(b)$ and $(c)$, defined in Section \ref{Sec:separation}.  

If we could ignore the hyper-cubic constraint sector $(c)$, we would find that the corrections coming from the remaining area-lengths constraints $(b)$ are of the same order $\Lambda^6/\lambda^2$, as found for the Area Regge action. The only change for the correction  $\Gamma_{(a_1)(a_1)}$ consists of replacing $H^{-1}_{(b'_1)(b'_1)}$ in (\ref{Hbs}) with
\ba
(H^{-1}_{(b'_1)(b'_1)})^\gamma= 
-\imath \frac{\gamma \lambda^2}{63488 }
\left(
\begin{array}{cccccc}
 \frac{5001}{71} & 27 & 27 & 27 & 27 & \frac{537}{71} \\
 27 & \frac{5001}{71} & 27 & 27 & \frac{537}{71} & 27 \\
 27 & 27 & \frac{5001}{71} & \frac{537}{71} & 27 & 27 \\
 27 & 27 & \frac{537}{71} & \frac{5001}{71} & 27 & 27 \\
 27 & \frac{537}{71} & 27 & 27 & \frac{5001}{71} & 27 \\
 \frac{537}{71} & 27 & 27 & 27 & 27 & \frac{5001}{71} \\
\end{array}
\right)  + {\cal O}(\gamma^2)   \q .
\ea

The issue is, however, that we cannot ignore the hyper-cubic constraint sector $(c)$. The reason is the following: 22 of the 23 eigenvalues for the $\Lambda^0$ part of $G_{(c)(c)}$ are of order $s^{-2}$. The one remaining eigenvalue is vanishing (to order $\Lambda^0$).  The block $G_{(b)(c)}$ does include terms of order $s^{-1}$. The corrections resulting from integrating out the hyper-cubical constraints to the $(b)(b)$ block could therefore be of order $s^0$ and also --- due to the one vanishing eigenvalue for the $\Lambda^0 s^{-2}$ part of $G$ --- include terms with $s^{-1}$ scaling.

If we aim at determining terms to order $s^0$ we have therefore to include terms of order $s^1$ into our calculations.  This makes the analysis much more involved, and might require a redefinition (in the $s^1$ order) of the hyper-cubic constraints, as compared to the version used in the current work.

\section{Discussion}\label{Disc}

The Area Regge action plays a key role for  spin foam models \cite{Perez}. The dynamics and the degrees of freedom encoded by this action are however not very well understood. In particular, no systematiic analysis of the continuum limit has been performed so far.

In this work we provided the first  analysis of  the linearized Area-Regge action for a triangulation based on a hyper-cubical lattice. In order to be able to define the Area-Regge action on this lattice we introduced a tilting with parameter $s$. We then considered the limit $s\rightarrow 0$ and the continuum limit. We found an effective action for the length fluctuation that agrees with a discretization of the linearized Einstein-Hilbert action, up to terms of ${\cal O}(s^1)$ order and terms of ${\cal O}(\lambda^4)$ order, where $\lambda^4$  is the lattice constant of the background lattice. We computed the ${\cal O}(\lambda^4)$ term in (\ref{FF}) and identified it to be quadratic in derivatives of curvature. 

This result has important implications for spin foams. The semi-classical limit of spin foam amplitudes is described by the Area Regge action. Area-Length constraints may or may not be included in the various models. They are in particular not included into the Barrett-Crane model. As the Area-Regge action demands on the discrete level vanishing deficit angles, it was concluded that this model is not suitable for a description of (quantized) gravity. But we find here that the continuum limit of the linearized theory (on the hyper-cubic lattice) does lead to the continuum Einstein-Hilbert action. The Barrett-Crane model might therefore still work as a model for quantum gravity. To gain more clarity on this question, we should however also consider lattices different from the hyper-cubic ones.

The spin foam amplitudes are approximated very well by their semi-classical limit for spins leading to  lattice constants $\lambda > 10^1 \ell_{\rm P}$. The areas in spin foams are actually discrete \cite{DiscreteGeom1}, the spectrum is (for larger areas) approximated\footnote{The area spectra for the Euclidean signature EPRL-FK model and the Barrett-Crane model do {\it not} include the Barbero--Immirzi parameter $\gamma$. This parameter does appear only for space-like areas in the Lorentzian signature EPRL-FK models.} by $A_t \sim j \ell_P^2$ with $j$ a half-integer (if one uses $\text{SU}(2) \times \text{SU}(2)$ as gauge group) or integer (if one uses $\text{SO}(4)$ as gauge group). It might therefore not be possible to set the tilting parameter $s$ exactly to zero. This area discreteness induced deviation of $s$ from zero is of the order $s\sim \ell_P^2/\lambda$, that is, with $\lambda > 10^1 \ell_{\rm P}$, a magnitude or more smaller than the Planck length.  It could however happen that the $s^1$--corrections are of lower order in the lattice constant than the $s^0$ corrections. We will consider these $s^1$--corrections in future work.

In order to find the continuum limit for Area Regge caclulus it was essential to separate the degrees of freedom into metric variables, area-length constraints, as well as spurious variables. The area-length constraints could be furthermore divided into the set of hyper-cubic constraints and the remaining area-length constraints, which here we called $\zeta$-variables. The order of the correction resulted then from some rather surprising scaling properties of the various Area Regge Hessian blocks associated to this variable splitting. That is the dynamics of Area Regge calculus itself leads to a continuum limit, which coincided with the continuum limit of Length Regge calculus.

This result is quite surprising, as the equations of motions for the Area Regge action do impose vanishing curvature. The equations of motions can however be split into different parts: one part includes only the length variables and does give the graviton equations of motion plus higher order corrections. Another part does  involve the area-length constraints but equates these constraints to certain functions of the length variables. These functions are of higher order in the lattice constant or of higher order in the tilting parameter. Sending both parameters to zero we thus impose the area-length constraints.

The limit $s\rightarrow 0$ seems to play a crucial role: it does implement a subset of the area-length constraints, which we named hyper-cubic constraints. 
To see whether the result depends on this feature, which might be specific to our choice of lattice, we should understand the continuum limit for other types of lattices, which might not feature hyper-cubic constraints. Note that a hyper-cube can be triangulated in a number of different ways \cite{Huggins}, this can lead to different classes of hyper-cubic lattices.\footnote{These different triangulations may however not allow to glue opposite faces of the hyper-cube to each other. One would have to combine hyper-cubes to a larger fundamental hyper-cubic cell, whose opposite faces can be glued to a lattice.}

It would be also useful to have a better understanding of the degrees of freedom in Area Regge calculus in the continuum limit, e.g. a match to torsion. This might allow to postulate a continuum version for the Area Regge action. It would then be much easier  to study the dynamics of this continuum version, as compared to the lattice version. 

We have also given an outlook on including area-length constraint terms into the action. The correction term does then scale with the Barbero--Immirzi parameter. The  corrections induced by integrating out the $\zeta$--variables are of the same general form as in the case of Area Regge calculus without the addition of these constraint terms. But the constraint terms do have a different scaling in $s$ for the blocks involving the hyper-cubic constraints. This seems to demand an explicit calculation of the $s^1$--order terms. This will be the subject of future work.

Here we have analyzed the continuum limit of the Area Regge action on the hyper-cubic lattice, for a Euclidean flat background. It would be of course highly interesting to consider also the Lorentzian case. Here we have a lattice constant for the time-like edges as well as a lattice constant for the space-like edges. We can adjust this ratio so that either the hyper-diagonals, the body-diagonals or the face-diagonals or neither become null in the background geometry.   Null edges play a special role in Regge calculus \cite{Nullstruts}, it would therefore be interesting to see whether the results depend on this choice. 

Another possible generalization concerns the Area Regge action. Instead of working with ($\text{SO}(4)$)--invariant variables on the lattice, we can go back to using the $B$--fields and constraints of the Plebanski action, from which spin foams are derived. This does introduce much more fields but could simplify the match between continuum and lattice quantities. It might than be possible to obtain effective lattice actions which in form reproduce modified BF-actions proposed by Krasnov in \cite{Krasnov}.

The perturbative expansion of Area Regge calculus, which we introduced here, can be also helpful to determine the measure terms in the (perturbative and non-perturbative) path integral and in particular for the effective spin foams \cite{EffSF1,EffSF2,EffSF3}.  Demanding triangulation invariance to one-loop order  the work \cite{PImeasure1} derives such measure terms for 3D Length Regge calculus.  This allowed the computation of physically interesting partition functions to one-loop, which established a holographic relation for non-asymptotic boundary in 3D quantum gravity  \cite{Hol3D4D}. These one-loop results have then been confirmed in the fully non-perturbative set-up in \cite{PRHolo}. For 4D it was found that Length Regge calculus does not admit a measure, which is invariant to one-loop order under changes of triangulation that leave the action invariant \cite{PImeasure2}. This could however be different for the Area Regge action.  The choice of measure plays an important role for the divergence structure of the spin foam path integral \cite{SFDiv} and are crucial for  the restoration of diffeomorphism symmetry in the continuum limit \cite{DiffDiv}.

\appendix
\section{Bases for the length, the $\zeta$- and the hyper-cubic constraint sector }\label{App:Basis}

Here we will give sets of basis vectors for the various sectors described in Section \ref{Sec:Decoupling}. We will use the variables $\alpha^{\pm}_{XX'}$ defined in Equation \ref{trafo2}. We order these variables according to the following list of indices $(X,X')^\pm$:
\ba\label{Eq:Basis}
&&(1,2)^+,(1,4)^+,(1,8)^+,(2,4)^+,(2,8)^+,(4,8)^+, \nn\\
&&(1,6)^+,(1,10)^+,(1,12)^+,(2,5)^+,(2,9)^+,(2,12)^+,(4,3)^+,(4,9)^+,(4,10)^+,(8,3)^+,(8,5)^+,(8,6)^+, \nn\\
&&(3,12)^+,(5,10)^+,(9,6)^+, \nn\\
&& \nn\\
&&(1,2)^-,(1,4)^-,(1,8)^-,(2,4)^-,(2,8)^-,(4,8)^-, \nn\\
&&(1,6)^-,(1,10)^-,(1,12)^-,(2,5)^-,(2,9)^-,(2,12)^-,(4,3)^-,(4,9)^-,(4,10)^-,(8,3)^-,(8,5)^-,(8,6)^-,\nn\\
&&(3,12)^-,(5,10)^-,(9,6)^-, \nn\\
&& \nn\\
&&(1,14)^+,(2,13)^+,(4,11)^+,(8,7)^+ ,\nn\\
&&(1,14)^-,(2,13)^-,(4,11)^-,(8,7)^- \q .
\ea
That is, we have first 21 $\alpha^+$--variables, then 21 $\alpha^-$--variables. The last eight variables are given by the 4 spurious $\alpha^+$--variables and the 4 spurious $\alpha^-$--variables.

The hyper-cubic constraints are given by the left null vectors of the Jacobians giving the derivatives of the area squares with respect to the lengths squares $(\partial A^2_t/\partial L^2_e)_\sigma$ for each of the 24 simplices of the hyper-cubes. 
Using the variables (\ref{trafo2}) in the ordering (\ref{Eq:Basis}) these null vectors can be expressed without any momentum dependence and are given by the columns of the following matrix:
\ba\label{appHCC1}
M_{\rm HCC}= 
{\tiny
\left(
\begin{array}{cccccccccccccccccccccccc}
 0 & 0 & 0 & 0 & 0 & 0 & 0 & 0 & 0 & 0 & 0 & 0 & 4 & 0 & 4 & 0 & 0 & 0 & 4 & 0 & 4 & 0 & 0 & 0 \\
 0 & 0 & 0 & 0 & 0 & 0 & 4 & 0 & 4 & 0 & 0 & 0 & 0 & 0 & 0 & 0 & 0 & 0 & 0 & 4 & 0 & 0 & 4 & 0 \\
 0 & 0 & 0 & 0 & 0 & 0 & 0 & 4 & 0 & 0 & 4 & 0 & 0 & 4 & 0 & 0 & 4 & 0 & 0 & 0 & 0 & 0 & 0 & 0 \\
 4 & 0 & 4 & 0 & 0 & 0 & 0 & 0 & 0 & 0 & 0 & 0 & 0 & 0 & 0 & 0 & 0 & 0 & 0 & 0 & 0 & 4 & 0 & 4 \\
 0 & 4 & 0 & 0 & 4 & 0 & 0 & 0 & 0 & 0 & 0 & 0 & 0 & 0 & 0 & 4 & 0 & 4 & 0 & 0 & 0 & 0 & 0 & 0 \\
 0 & 0 & 0 & 4 & 0 & 4 & 0 & 0 & 0 & 4 & 0 & 4 & 0 & 0 & 0 & 0 & 0 & 0 & 0 & 0 & 0 & 0 & 0 & 0 \\
 0 & 0 & 0 & 0 & 0 & 0 & 0 & 0 & -2 & 0 & 0 & 0 & 0 & 0 & -2 & 0 & 0 & 0 & -2 & -2 & 0 & 0 & 0 & 0 \\
 0 & 0 & 0 & 0 & 0 & 0 & 0 & 0 & 0 & 0 & -2 & 0 & -2 & -2 & 0 & 0 & 0 & 0 & 0 & 0 & -2 & 0 & 0 & 0 \\
 0 & 0 & 0 & 0 & 0 & 0 & -2 & -2 & 0 & 0 & 0 & 0 & 0 & 0 & 0 & 0 & -2 & 0 & 0 & 0 & 0 & 0 & -2 & 0 \\
 0 & 0 & -2 & 0 & 0 & 0 & 0 & 0 & 0 & 0 & 0 & 0 & -2 & 0 & 0 & 0 & 0 & 0 & 0 & 0 & -2 & -2 & 0 & 0 \\
 0 & 0 & 0 & 0 & -2 & 0 & 0 & 0 & 0 & 0 & 0 & 0 & 0 & 0 & -2 & -2 & 0 & 0 & -2 & 0 & 0 & 0 & 0 & 0 \\
 -2 & -2 & 0 & 0 & 0 & 0 & 0 & 0 & 0 & 0 & 0 & 0 & 0 & 0 & 0 & 0 & 0 & -2 & 0 & 0 & 0 & 0 & 0 & -2 \\
 -2 & 0 & 0 & 0 & 0 & 0 & -2 & 0 & 0 & 0 & 0 & 0 & 0 & 0 & 0 & 0 & 0 & 0 & 0 & 0 & 0 & 0 & -2 & -2 \\
 0 & 0 & 0 & 0 & 0 & -2 & 0 & 0 & -2 & -2 & 0 & 0 & 0 & 0 & 0 & 0 & 0 & 0 & 0 & -2 & 0 & 0 & 0 & 0 \\
 0 & 0 & -2 & -2 & 0 & 0 & 0 & 0 & 0 & 0 & 0 & -2 & 0 & 0 & 0 & 0 & 0 & 0 & 0 & 0 & 0 & -2 & 0 & 0 \\
 0 & -2 & 0 & 0 & 0 & 0 & 0 & -2 & 0 & 0 & 0 & 0 & 0 & 0 & 0 & 0 & -2 & -2 & 0 & 0 & 0 & 0 & 0 & 0 \\
 0 & 0 & 0 & -2 & 0 & 0 & 0 & 0 & 0 & 0 & -2 & -2 & 0 & -2 & 0 & 0 & 0 & 0 & 0 & 0 & 0 & 0 & 0 & 0 \\
 0 & 0 & 0 & 0 & -2 & -2 & 0 & 0 & 0 & -2 & 0 & 0 & 0 & 0 & 0 & -2 & 0 & 0 & 0 & 0 & 0 & 0 & 0 & 0 \\
 1 & 1 & 0 & 0 & 0 & 0 & 1 & 1 & 0 & 0 & 0 & 0 & 0 & 0 & 0 & 0 & 1 & 1 & 0 & 0 & 0 & 0 & 1 & 1 \\
 0 & 0 & 1 & 1 & 0 & 0 & 0 & 0 & 0 & 0 & 1 & 1 & 1 & 1 & 0 & 0 & 0 & 0 & 0 & 0 & 1 & 1 & 0 & 0 \\
 0 & 0 & 0 & 0 & 1 & 1 & 0 & 0 & 1 & 1 & 0 & 0 & 0 & 0 & 1 & 1 & 0 & 0 & 1 & 1 & 0 & 0 & 0 & 0 \\
 0 & 0 & 0 & 0 & 0 & 0 & 0 & 0 & 0 & 0 & 0 & 0 & 4 & 0 & -4 & 0 & 0 & 0 & 4 & 0 & -4 & 0 & 0 & 0 \\
 0 & 0 & 0 & 0 & 0 & 0 & 4 & 0 & -4 & 0 & 0 & 0 & 0 & 0 & 0 & 0 & 0 & 0 & 0 & 4 & 0 & 0 & -4 & 0 \\
 0 & 0 & 0 & 0 & 0 & 0 & 0 & 4 & 0 & 0 & -4 & 0 & 0 & 4 & 0 & 0 & -4 & 0 & 0 & 0 & 0 & 0 & 0 & 0 \\
 4 & 0 & -4 & 0 & 0 & 0 & 0 & 0 & 0 & 0 & 0 & 0 & 0 & 0 & 0 & 0 & 0 & 0 & 0 & 0 & 0 & 4 & 0 & -4 \\
 0 & 4 & 0 & 0 & -4 & 0 & 0 & 0 & 0 & 0 & 0 & 0 & 0 & 0 & 0 & 4 & 0 & -4 & 0 & 0 & 0 & 0 & 0 & 0 \\
 0 & 0 & 0 & 4 & 0 & -4 & 0 & 0 & 0 & 4 & 0 & -4 & 0 & 0 & 0 & 0 & 0 & 0 & 0 & 0 & 0 & 0 & 0 & 0 \\
 0 & 0 & 0 & 0 & 0 & 0 & 0 & 0 & 2 & 0 & 0 & 0 & 0 & 0 & 2 & 0 & 0 & 0 & -2 & -2 & 0 & 0 & 0 & 0 \\
 0 & 0 & 0 & 0 & 0 & 0 & 0 & 0 & 0 & 0 & 2 & 0 & -2 & -2 & 0 & 0 & 0 & 0 & 0 & 0 & 2 & 0 & 0 & 0 \\
 0 & 0 & 0 & 0 & 0 & 0 & -2 & -2 & 0 & 0 & 0 & 0 & 0 & 0 & 0 & 0 & 2 & 0 & 0 & 0 & 0 & 0 & 2 & 0 \\
 0 & 0 & 2 & 0 & 0 & 0 & 0 & 0 & 0 & 0 & 0 & 0 & 2 & 0 & 0 & 0 & 0 & 0 & 0 & 0 & -2 & -2 & 0 & 0 \\
 0 & 0 & 0 & 0 & 2 & 0 & 0 & 0 & 0 & 0 & 0 & 0 & 0 & 0 & -2 & -2 & 0 & 0 & 2 & 0 & 0 & 0 & 0 & 0 \\
 -2 & -2 & 0 & 0 & 0 & 0 & 0 & 0 & 0 & 0 & 0 & 0 & 0 & 0 & 0 & 0 & 0 & 2 & 0 & 0 & 0 & 0 & 0 & 2 \\
 2 & 0 & 0 & 0 & 0 & 0 & 2 & 0 & 0 & 0 & 0 & 0 & 0 & 0 & 0 & 0 & 0 & 0 & 0 & 0 & 0 & 0 & -2 & -2 \\
 0 & 0 & 0 & 0 & 0 & 2 & 0 & 0 & -2 & -2 & 0 & 0 & 0 & 0 & 0 & 0 & 0 & 0 & 0 & 2 & 0 & 0 & 0 & 0 \\
 0 & 0 & -2 & -2 & 0 & 0 & 0 & 0 & 0 & 0 & 0 & 2 & 0 & 0 & 0 & 0 & 0 & 0 & 0 & 0 & 0 & 2 & 0 & 0 \\
 0 & 2 & 0 & 0 & 0 & 0 & 0 & 2 & 0 & 0 & 0 & 0 & 0 & 0 & 0 & 0 & -2 & -2 & 0 & 0 & 0 & 0 & 0 & 0 \\
 0 & 0 & 0 & 2 & 0 & 0 & 0 & 0 & 0 & 0 & -2 & -2 & 0 & 2 & 0 & 0 & 0 & 0 & 0 & 0 & 0 & 0 & 0 & 0 \\
 0 & 0 & 0 & 0 & -2 & -2 & 0 & 0 & 0 & 2 & 0 & 0 & 0 & 0 & 0 & 2 & 0 & 0 & 0 & 0 & 0 & 0 & 0 & 0 \\
 1 & 1 & 0 & 0 & 0 & 0 & 1 & 1 & 0 & 0 & 0 & 0 & 0 & 0 & 0 & 0 & -1 & -1 & 0 & 0 & 0 & 0 & -1 & -1 \\
 0 & 0 & 1 & 1 & 0 & 0 & 0 & 0 & 0 & 0 & -1 & -1 & 1 & 1 & 0 & 0 & 0 & 0 & 0 & 0 & -1 & -1 & 0 & 0 \\
 0 & 0 & 0 & 0 & 1 & 1 & 0 & 0 & -1 & -1 & 0 & 0 & 0 & 0 & -1 & -1 & 0 & 0 & 1 & 1 & 0 & 0 & 0 & 0 \\
 0 & 0 & 0 & 0 & 0 & 0 & 0 & 0 & 0 & 0 & 0 & 0 & 0 & 0 & 0 & 0 & 0 & 0 & 0 & 0 & 0 & 0 & 0 & 0 \\
 0 & 0 & 0 & 0 & 0 & 0 & 0 & 0 & 0 & 0 & 0 & 0 & 0 & 0 & 0 & 0 & 0 & 0 & 0 & 0 & 0 & 0 & 0 & 0 \\
 0 & 0 & 0 & 0 & 0 & 0 & 0 & 0 & 0 & 0 & 0 & 0 & 0 & 0 & 0 & 0 & 0 & 0 & 0 & 0 & 0 & 0 & 0 & 0 \\
 0 & 0 & 0 & 0 & 0 & 0 & 0 & 0 & 0 & 0 & 0 & 0 & 0 & 0 & 0 & 0 & 0 & 0 & 0 & 0 & 0 & 0 & 0 & 0 \\
 0 & 0 & 0 & 0 & 0 & 0 & 0 & 0 & 0 & 0 & 0 & 0 & 0 & 0 & 0 & 0 & 0 & 0 & 0 & 0 & 0 & 0 & 0 & 0 \\
 0 & 0 & 0 & 0 & 0 & 0 & 0 & 0 & 0 & 0 & 0 & 0 & 0 & 0 & 0 & 0 & 0 & 0 & 0 & 0 & 0 & 0 & 0 & 0 \\
 0 & 0 & 0 & 0 & 0 & 0 & 0 & 0 & 0 & 0 & 0 & 0 & 0 & 0 & 0 & 0 & 0 & 0 & 0 & 0 & 0 & 0 & 0 & 0 \\
 0 & 0 & 0 & 0 & 0 & 0 & 0 & 0 & 0 & 0 & 0 & 0 & 0 & 0 & 0 & 0 & 0 & 0 & 0 & 0 & 0 & 0 & 0 & 0 \\
\end{array}
\right)
}\, .
\ea
We discussed that the hyper-cubic constraints separate the $\alpha^+$ and the $\alpha^-$ sector, but this is not obvious with the form of the constraints in (\ref{appHCC1}). It will become evident if we construct an orthogonal projector $P'_{\rm HCC}$ onto the constraints. To this end we first note that the 24 constraints in (\ref{appHCC1}) are not independent.   $M_{\rm HCC}$ has precisely one right null vector given by
\ba
n^\dagger_{\rm HCC}=
\left(
\begin{array}{cccccccccccccccccccccccc}
 1 & -1 & -1 & 1 & 1 & -1 & -1 & 1 & 1 & -1 & -1 & 1 & 1 & -1 & -1 & 1 & 1 & -1 & -1 & 1 & 1 & -1 & -1 & 1 \\
\end{array}
\right)\q .
\ea
This null vector will also appear for the Gram matrix, formed from the inner products between the constraint vectors, that is the columns of $M_{\rm HCC}$:
\ba
{\cal G}_{\rm HCC} = (M_{HCC})^\dagger \cdot M_{HCC}      \q .
\ea
In order to invert this Gram matrix we add a term $\beta n_{\rm HCC} n^\dagger_{\rm HCC}$ to it. The orthogonal projector onto the hyper-cubic constraint sector can then be defined as
\ba
P'_{\rm HCC} =   M_{HCC} \,\cdot \,  (   {\cal G}_{\rm HCC} + \beta    n_{\rm HCC} n^\dagger_{\rm HCC})^{-1}  \,  \cdot \,    (M_{HCC})^\dagger \q ,
\ea
and is independent of $\beta$. This projector has the following block form (where the sub-indices indicate the size of the blocks)
\ba\label{PHCC}
P'_{HCC}=
\left( 
\begin{array}{ccc}
C^+_{21\times 21}&0_{21 \times 21}&0_{21\times 8} \\
0_{21 \times 21}&C^+_{21\times 21}&0_{21\times 8} \\
0_{8\times 21}&0_{8\times 21}&0_{8\times 8}
\end{array}
\right) \q ,
\ea
with the blocks $C^+_{21\times 21}$ and $C^-_{21\times 21}$ given by
\ba
C^+_{21\times 21}= \hspace{17cm}
\nn\\
{\tiny\left(
\begin{array}{ccccccccccccccccccccc}
 \frac{163}{195} & -\frac{5}{78} & -\frac{5}{78} & -\frac{5}{78} & -\frac{5}{78} & \frac{7}{195} & -\frac{59}{390} & -\frac{59}{390} & -\frac{2}{39}
& -\frac{59}{390} & -\frac{59}{390} & -\frac{2}{39} & -\frac{2}{39} & \frac{19}{390} & \frac{19}{390} & -\frac{2}{39} & \frac{19}{390} & \frac{19}{390}
& \frac{2}{39} & \frac{2}{39} & \frac{2}{39} \\
 -\frac{5}{78} & \frac{163}{195} & -\frac{5}{78} & -\frac{5}{78} & \frac{7}{195} & -\frac{5}{78} & -\frac{59}{390} & -\frac{2}{39} & -\frac{59}{390}
& -\frac{2}{39} & \frac{19}{390} & \frac{19}{390} & -\frac{59}{390} & -\frac{59}{390} & -\frac{2}{39} & \frac{19}{390} & -\frac{2}{39} & \frac{19}{390}
& \frac{2}{39} & \frac{2}{39} & \frac{2}{39} \\
 -\frac{5}{78} & -\frac{5}{78} & \frac{163}{195} & \frac{7}{195} & -\frac{5}{78} & -\frac{5}{78} & -\frac{2}{39} & -\frac{59}{390} & -\frac{59}{390}
& \frac{19}{390} & -\frac{2}{39} & \frac{19}{390} & \frac{19}{390} & -\frac{2}{39} & \frac{19}{390} & -\frac{59}{390} & -\frac{59}{390} & -\frac{2}{39}
& \frac{2}{39} & \frac{2}{39} & \frac{2}{39} \\
 -\frac{5}{78} & -\frac{5}{78} & \frac{7}{195} & \frac{163}{195} & -\frac{5}{78} & -\frac{5}{78} & -\frac{2}{39} & \frac{19}{390} & \frac{19}{390}
& -\frac{59}{390} & -\frac{2}{39} & -\frac{59}{390} & -\frac{59}{390} & -\frac{2}{39} & -\frac{59}{390} & \frac{19}{390} & \frac{19}{390} & -\frac{2}{39}
& \frac{2}{39} & \frac{2}{39} & \frac{2}{39} \\
 -\frac{5}{78} & \frac{7}{195} & -\frac{5}{78} & -\frac{5}{78} & \frac{163}{195} & -\frac{5}{78} & \frac{19}{390} & -\frac{2}{39} & \frac{19}{390}
& -\frac{2}{39} & -\frac{59}{390} & -\frac{59}{390} & \frac{19}{390} & \frac{19}{390} & -\frac{2}{39} & -\frac{59}{390} & -\frac{2}{39} & -\frac{59}{390}
& \frac{2}{39} & \frac{2}{39} & \frac{2}{39} \\
 \frac{7}{195} & -\frac{5}{78} & -\frac{5}{78} & -\frac{5}{78} & -\frac{5}{78} & \frac{163}{195} & \frac{19}{390} & \frac{19}{390} & -\frac{2}{39}
& \frac{19}{390} & \frac{19}{390} & -\frac{2}{39} & -\frac{2}{39} & -\frac{59}{390} & -\frac{59}{390} & -\frac{2}{39} & -\frac{59}{390} & -\frac{59}{390}
& \frac{2}{39} & \frac{2}{39} & \frac{2}{39} \\
 -\frac{59}{390} & -\frac{59}{390} & -\frac{2}{39} & -\frac{2}{39} & \frac{19}{390} & \frac{19}{390} & \frac{179}{390} & -\frac{11}{78} & -\frac{11}{78}
& -\frac{11}{78} & \frac{31}{195} & \frac{23}{390} & -\frac{11}{78} & \frac{31}{195} & \frac{23}{390} & \frac{23}{390} & \frac{23}{390} & -\frac{11}{78}
& \frac{8}{195} & \frac{8}{195} & -\frac{31}{195} \\
 -\frac{59}{390} & -\frac{2}{39} & -\frac{59}{390} & \frac{19}{390} & -\frac{2}{39} & \frac{19}{390} & -\frac{11}{78} & \frac{179}{390} & -\frac{11}{78}
& \frac{31}{195} & -\frac{11}{78} & \frac{23}{390} & \frac{23}{390} & \frac{23}{390} & -\frac{11}{78} & -\frac{11}{78} & \frac{31}{195} & \frac{23}{390}
& \frac{8}{195} & -\frac{31}{195} & \frac{8}{195} \\
 -\frac{2}{39} & -\frac{59}{390} & -\frac{59}{390} & \frac{19}{390} & \frac{19}{390} & -\frac{2}{39} & -\frac{11}{78} & -\frac{11}{78} & \frac{179}{390}
& \frac{23}{390} & \frac{23}{390} & -\frac{11}{78} & \frac{31}{195} & -\frac{11}{78} & \frac{23}{390} & \frac{31}{195} & -\frac{11}{78} & \frac{23}{390}
& -\frac{31}{195} & \frac{8}{195} & \frac{8}{195} \\
 -\frac{59}{390} & -\frac{2}{39} & \frac{19}{390} & -\frac{59}{390} & -\frac{2}{39} & \frac{19}{390} & -\frac{11}{78} & \frac{31}{195} & \frac{23}{390}
& \frac{179}{390} & -\frac{11}{78} & -\frac{11}{78} & -\frac{11}{78} & \frac{23}{390} & \frac{31}{195} & \frac{23}{390} & -\frac{11}{78} & \frac{23}{390}
& \frac{8}{195} & -\frac{31}{195} & \frac{8}{195} \\
 -\frac{59}{390} & \frac{19}{390} & -\frac{2}{39} & -\frac{2}{39} & -\frac{59}{390} & \frac{19}{390} & \frac{31}{195} & -\frac{11}{78} & \frac{23}{390}
& -\frac{11}{78} & \frac{179}{390} & -\frac{11}{78} & \frac{23}{390} & -\frac{11}{78} & \frac{23}{390} & -\frac{11}{78} & \frac{23}{390} & \frac{31}{195}
& \frac{8}{195} & \frac{8}{195} & -\frac{31}{195} \\
 -\frac{2}{39} & \frac{19}{390} & \frac{19}{390} & -\frac{59}{390} & -\frac{59}{390} & -\frac{2}{39} & \frac{23}{390} & \frac{23}{390} & -\frac{11}{78}
& -\frac{11}{78} & -\frac{11}{78} & \frac{179}{390} & \frac{31}{195} & \frac{23}{390} & -\frac{11}{78} & \frac{31}{195} & \frac{23}{390} & -\frac{11}{78}
& -\frac{31}{195} & \frac{8}{195} & \frac{8}{195} \\
 -\frac{2}{39} & -\frac{59}{390} & \frac{19}{390} & -\frac{59}{390} & \frac{19}{390} & -\frac{2}{39} & -\frac{11}{78} & \frac{23}{390} & \frac{31}{195}
& -\frac{11}{78} & \frac{23}{390} & \frac{31}{195} & \frac{179}{390} & -\frac{11}{78} & -\frac{11}{78} & -\frac{11}{78} & \frac{23}{390} & \frac{23}{390}
& -\frac{31}{195} & \frac{8}{195} & \frac{8}{195} \\
 \frac{19}{390} & -\frac{59}{390} & -\frac{2}{39} & -\frac{2}{39} & \frac{19}{390} & -\frac{59}{390} & \frac{31}{195} & \frac{23}{390} & -\frac{11}{78}
& \frac{23}{390} & -\frac{11}{78} & \frac{23}{390} & -\frac{11}{78} & \frac{179}{390} & -\frac{11}{78} & \frac{23}{390} & -\frac{11}{78} & \frac{31}{195}
& \frac{8}{195} & \frac{8}{195} & -\frac{31}{195} \\
 \frac{19}{390} & -\frac{2}{39} & \frac{19}{390} & -\frac{59}{390} & -\frac{2}{39} & -\frac{59}{390} & \frac{23}{390} & -\frac{11}{78} & \frac{23}{390}
& \frac{31}{195} & \frac{23}{390} & -\frac{11}{78} & -\frac{11}{78} & -\frac{11}{78} & \frac{179}{390} & \frac{23}{390} & \frac{31}{195} & -\frac{11}{78}
& \frac{8}{195} & -\frac{31}{195} & \frac{8}{195} \\
 -\frac{2}{39} & \frac{19}{390} & -\frac{59}{390} & \frac{19}{390} & -\frac{59}{390} & -\frac{2}{39} & \frac{23}{390} & -\frac{11}{78} & \frac{31}{195}
& \frac{23}{390} & -\frac{11}{78} & \frac{31}{195} & -\frac{11}{78} & \frac{23}{390} & \frac{23}{390} & \frac{179}{390} & -\frac{11}{78} & -\frac{11}{78}
& -\frac{31}{195} & \frac{8}{195} & \frac{8}{195} \\
 \frac{19}{390} & -\frac{2}{39} & -\frac{59}{390} & \frac{19}{390} & -\frac{2}{39} & -\frac{59}{390} & \frac{23}{390} & \frac{31}{195} & -\frac{11}{78}
& -\frac{11}{78} & \frac{23}{390} & \frac{23}{390} & \frac{23}{390} & -\frac{11}{78} & \frac{31}{195} & -\frac{11}{78} & \frac{179}{390} & -\frac{11}{78}
& \frac{8}{195} & -\frac{31}{195} & \frac{8}{195} \\
 \frac{19}{390} & \frac{19}{390} & -\frac{2}{39} & -\frac{2}{39} & -\frac{59}{390} & -\frac{59}{390} & -\frac{11}{78} & \frac{23}{390} & \frac{23}{390}
& \frac{23}{390} & \frac{31}{195} & -\frac{11}{78} & \frac{23}{390} & \frac{31}{195} & -\frac{11}{78} & -\frac{11}{78} & -\frac{11}{78} & \frac{179}{390}
& \frac{8}{195} & \frac{8}{195} & -\frac{31}{195} \\
 \frac{2}{39} & \frac{2}{39} & \frac{2}{39} & \frac{2}{39} & \frac{2}{39} & \frac{2}{39} & \frac{8}{195} & \frac{8}{195} & -\frac{31}{195} & \frac{8}{195}
& \frac{8}{195} & -\frac{31}{195} & -\frac{31}{195} & \frac{8}{195} & \frac{8}{195} & -\frac{31}{195} & \frac{8}{195} & \frac{8}{195} & \frac{31}{195}
& -\frac{8}{195} & -\frac{8}{195} \\
 \frac{2}{39} & \frac{2}{39} & \frac{2}{39} & \frac{2}{39} & \frac{2}{39} & \frac{2}{39} & \frac{8}{195} & -\frac{31}{195} & \frac{8}{195} & -\frac{31}{195}
& \frac{8}{195} & \frac{8}{195} & \frac{8}{195} & \frac{8}{195} & -\frac{31}{195} & \frac{8}{195} & -\frac{31}{195} & \frac{8}{195} & -\frac{8}{195}
& \frac{31}{195} & -\frac{8}{195} \\
 \frac{2}{39} & \frac{2}{39} & \frac{2}{39} & \frac{2}{39} & \frac{2}{39} & \frac{2}{39} & -\frac{31}{195} & \frac{8}{195} & \frac{8}{195} & \frac{8}{195}
& -\frac{31}{195} & \frac{8}{195} & \frac{8}{195} & -\frac{31}{195} & \frac{8}{195} & \frac{8}{195} & \frac{8}{195} & -\frac{31}{195} & -\frac{8}{195}
& -\frac{8}{195} & \frac{31}{195} \\
\end{array}
\right)
}\nn
\ea
and
\ba
C^-_{21\times 21}= \hspace{17cm} \nn\\
\tiny{
\left(
\begin{array}{ccccccccccccccccccccc}
 \frac{33}{38} & -\frac{5}{76} & -\frac{5}{76} & \frac{5}{76} & \frac{5}{76} & 0 & -\frac{9}{76} & -\frac{9}{76} & -\frac{5}{38} & \frac{9}{76} &
\frac{9}{76} & \frac{5}{38} & 0 & -\frac{1}{76} & \frac{1}{76} & 0 & -\frac{1}{76} & \frac{1}{76} & 0 & \frac{1}{19} & \frac{1}{19} \\
 -\frac{5}{76} & \frac{33}{38} & -\frac{5}{76} & -\frac{5}{76} & 0 & \frac{5}{76} & -\frac{9}{76} & -\frac{5}{38} & -\frac{9}{76} & 0 & -\frac{1}{76}
& \frac{1}{76} & \frac{9}{76} & \frac{9}{76} & \frac{5}{38} & -\frac{1}{76} & 0 & \frac{1}{76} & \frac{1}{19} & 0 & \frac{1}{19} \\
 -\frac{5}{76} & -\frac{5}{76} & \frac{33}{38} & 0 & -\frac{5}{76} & -\frac{5}{76} & -\frac{5}{38} & -\frac{9}{76} & -\frac{9}{76} & -\frac{1}{76}
& 0 & \frac{1}{76} & -\frac{1}{76} & 0 & \frac{1}{76} & \frac{9}{76} & \frac{9}{76} & \frac{5}{38} & \frac{1}{19} & \frac{1}{19} & 0 \\
 \frac{5}{76} & -\frac{5}{76} & 0 & \frac{33}{38} & -\frac{5}{76} & \frac{5}{76} & 0 & -\frac{1}{76} & \frac{1}{76} & -\frac{9}{76} & -\frac{5}{38}
& -\frac{9}{76} & \frac{9}{76} & \frac{5}{38} & \frac{9}{76} & -\frac{1}{76} & \frac{1}{76} & 0 & \frac{1}{19} & -\frac{1}{19} & 0 \\
 \frac{5}{76} & 0 & -\frac{5}{76} & -\frac{5}{76} & \frac{33}{38} & -\frac{5}{76} & -\frac{1}{76} & 0 & \frac{1}{76} & -\frac{5}{38} & -\frac{9}{76}
& -\frac{9}{76} & -\frac{1}{76} & \frac{1}{76} & 0 & \frac{9}{76} & \frac{5}{38} & \frac{9}{76} & \frac{1}{19} & 0 & -\frac{1}{19} \\
 0 & \frac{5}{76} & -\frac{5}{76} & \frac{5}{76} & -\frac{5}{76} & \frac{33}{38} & -\frac{1}{76} & \frac{1}{76} & 0 & -\frac{1}{76} & \frac{1}{76}
& 0 & -\frac{5}{38} & -\frac{9}{76} & -\frac{9}{76} & \frac{5}{38} & \frac{9}{76} & \frac{9}{76} & 0 & \frac{1}{19} & -\frac{1}{19} \\
 -\frac{9}{76} & -\frac{9}{76} & -\frac{5}{38} & 0 & -\frac{1}{76} & -\frac{1}{76} & \frac{10}{19} & -\frac{13}{76} & -\frac{13}{76} & \frac{3}{76}
& -\frac{3}{19} & \frac{5}{76} & \frac{3}{76} & -\frac{3}{19} & \frac{5}{76} & \frac{5}{76} & \frac{5}{76} & -\frac{4}{19} & \frac{1}{19} & \frac{1}{19}
& -\frac{3}{19} \\
 -\frac{9}{76} & -\frac{5}{38} & -\frac{9}{76} & -\frac{1}{76} & 0 & \frac{1}{76} & -\frac{13}{76} & \frac{10}{19} & -\frac{13}{76} & -\frac{3}{19}
& \frac{3}{76} & \frac{5}{76} & \frac{5}{76} & \frac{5}{76} & -\frac{4}{19} & \frac{3}{76} & -\frac{3}{19} & \frac{5}{76} & \frac{1}{19} & -\frac{3}{19}
& \frac{1}{19} \\
 -\frac{5}{38} & -\frac{9}{76} & -\frac{9}{76} & \frac{1}{76} & \frac{1}{76} & 0 & -\frac{13}{76} & -\frac{13}{76} & \frac{10}{19} & \frac{5}{76}
& \frac{5}{76} & -\frac{4}{19} & -\frac{3}{19} & \frac{3}{76} & \frac{5}{76} & -\frac{3}{19} & \frac{3}{76} & \frac{5}{76} & -\frac{3}{19} & \frac{1}{19}
& \frac{1}{19} \\
 \frac{9}{76} & 0 & -\frac{1}{76} & -\frac{9}{76} & -\frac{5}{38} & -\frac{1}{76} & \frac{3}{76} & -\frac{3}{19} & \frac{5}{76} & \frac{10}{19} &
-\frac{13}{76} & -\frac{13}{76} & \frac{3}{76} & \frac{5}{76} & -\frac{3}{19} & \frac{5}{76} & -\frac{4}{19} & \frac{5}{76} & \frac{1}{19} & \frac{3}{19}
& -\frac{1}{19} \\
 \frac{9}{76} & -\frac{1}{76} & 0 & -\frac{5}{38} & -\frac{9}{76} & \frac{1}{76} & -\frac{3}{19} & \frac{3}{76} & \frac{5}{76} & -\frac{13}{76} &
\frac{10}{19} & -\frac{13}{76} & \frac{5}{76} & -\frac{4}{19} & \frac{5}{76} & \frac{3}{76} & \frac{5}{76} & -\frac{3}{19} & \frac{1}{19} & -\frac{1}{19}
& \frac{3}{19} \\
 \frac{5}{38} & \frac{1}{76} & \frac{1}{76} & -\frac{9}{76} & -\frac{9}{76} & 0 & \frac{5}{76} & \frac{5}{76} & -\frac{4}{19} & -\frac{13}{76} &
-\frac{13}{76} & \frac{10}{19} & -\frac{3}{19} & \frac{5}{76} & \frac{3}{76} & -\frac{3}{19} & \frac{5}{76} & \frac{3}{76} & -\frac{3}{19} & -\frac{1}{19}
& -\frac{1}{19} \\
 0 & \frac{9}{76} & -\frac{1}{76} & \frac{9}{76} & -\frac{1}{76} & -\frac{5}{38} & \frac{3}{76} & \frac{5}{76} & -\frac{3}{19} & \frac{3}{76} & \frac{5}{76}
& -\frac{3}{19} & \frac{10}{19} & -\frac{13}{76} & -\frac{13}{76} & -\frac{4}{19} & \frac{5}{76} & \frac{5}{76} & \frac{3}{19} & \frac{1}{19} & -\frac{1}{19}
\\
 -\frac{1}{76} & \frac{9}{76} & 0 & \frac{5}{38} & \frac{1}{76} & -\frac{9}{76} & -\frac{3}{19} & \frac{5}{76} & \frac{3}{76} & \frac{5}{76} & -\frac{4}{19}
& \frac{5}{76} & -\frac{13}{76} & \frac{10}{19} & -\frac{13}{76} & \frac{5}{76} & \frac{3}{76} & -\frac{3}{19} & -\frac{1}{19} & \frac{1}{19} & \frac{3}{19}
\\
 \frac{1}{76} & \frac{5}{38} & \frac{1}{76} & \frac{9}{76} & 0 & -\frac{9}{76} & \frac{5}{76} & -\frac{4}{19} & \frac{5}{76} & -\frac{3}{19} & \frac{5}{76}
& \frac{3}{76} & -\frac{13}{76} & -\frac{13}{76} & \frac{10}{19} & \frac{5}{76} & -\frac{3}{19} & \frac{3}{76} & -\frac{1}{19} & -\frac{3}{19} &
-\frac{1}{19} \\
 0 & -\frac{1}{76} & \frac{9}{76} & -\frac{1}{76} & \frac{9}{76} & \frac{5}{38} & \frac{5}{76} & \frac{3}{76} & -\frac{3}{19} & \frac{5}{76} & \frac{3}{76}
& -\frac{3}{19} & -\frac{4}{19} & \frac{5}{76} & \frac{5}{76} & \frac{10}{19} & -\frac{13}{76} & -\frac{13}{76} & \frac{3}{19} & -\frac{1}{19} &
\frac{1}{19} \\
 -\frac{1}{76} & 0 & \frac{9}{76} & \frac{1}{76} & \frac{5}{38} & \frac{9}{76} & \frac{5}{76} & -\frac{3}{19} & \frac{3}{76} & -\frac{4}{19} & \frac{5}{76}
& \frac{5}{76} & \frac{5}{76} & \frac{3}{76} & -\frac{3}{19} & -\frac{13}{76} & \frac{10}{19} & -\frac{13}{76} & -\frac{1}{19} & \frac{3}{19} & \frac{1}{19}
\\
 \frac{1}{76} & \frac{1}{76} & \frac{5}{38} & 0 & \frac{9}{76} & \frac{9}{76} & -\frac{4}{19} & \frac{5}{76} & \frac{5}{76} & \frac{5}{76} & -\frac{3}{19}
& \frac{3}{76} & \frac{5}{76} & -\frac{3}{19} & \frac{3}{76} & -\frac{13}{76} & -\frac{13}{76} & \frac{10}{19} & -\frac{1}{19} & -\frac{1}{19} &
-\frac{3}{19} \\
 0 & \frac{1}{19} & \frac{1}{19} & \frac{1}{19} & \frac{1}{19} & 0 & \frac{1}{19} & \frac{1}{19} & -\frac{3}{19} & \frac{1}{19} & \frac{1}{19} &
-\frac{3}{19} & \frac{3}{19} & -\frac{1}{19} & -\frac{1}{19} & \frac{3}{19} & -\frac{1}{19} & -\frac{1}{19} & \frac{3}{19} & 0 & 0 \\
 \frac{1}{19} & 0 & \frac{1}{19} & -\frac{1}{19} & 0 & \frac{1}{19} & \frac{1}{19} & -\frac{3}{19} & \frac{1}{19} & \frac{3}{19} & -\frac{1}{19}
& -\frac{1}{19} & \frac{1}{19} & \frac{1}{19} & -\frac{3}{19} & -\frac{1}{19} & \frac{3}{19} & -\frac{1}{19} & 0 & \frac{3}{19} & 0 \\
 \frac{1}{19} & \frac{1}{19} & 0 & 0 & -\frac{1}{19} & -\frac{1}{19} & -\frac{3}{19} & \frac{1}{19} & \frac{1}{19} & -\frac{1}{19} & \frac{3}{19}
& -\frac{1}{19} & -\frac{1}{19} & \frac{3}{19} & -\frac{1}{19} & \frac{1}{19} & \frac{1}{19} & -\frac{3}{19} & 0 & 0 & \frac{3}{19} \\
\end{array}
\right)
}\; .
\q\q\q\q\q\q
\nn
\ea

The projector $P_{\rm HCC}=\mathbb{I}_{50}-P'_{\rm HCC}$ annihilates the hyper-cubic constraints.  The projector (\ref{PHCC}) makes the split of the constraints into the $\alpha^+$ and the $\alpha^-$ sector obvious. 
The rows (or columns) of $C^\pm_{21\times 21}$ define also the hyper-cubic constraints --- but with an over-complete set. There are only $11$ independent constraints $\chi^+_{c_+}$ and $12$ independent constraints $\chi^-_{c_-}$. 
The columns of the following matrix $M_{\chi}$ provide a basis for the $\chi^+$ and the $\chi^-$ constraints:
\ba
M_{\chi}=
{\scriptsize
\left(
\begin{array}{ccccccccccccccccccccccc}
 2 & -\frac{1}{4} & -\frac{1}{4} & -1 & -2 & -1 & 0 & 0 & 0 & 0 & 0 & 0 & 0 & 0 & 0 & 0 & 0 & 0 & 0 & 0 & 0 & 0 & 0 \\
 2 & -\frac{1}{4} & \frac{1}{2} & -1 & 0 & -1 & 0 & 0 & 0 & 0 & 0 & 0 & 0 & 0 & 0 & 0 & 0 & 0 & 0 & 0 & 0 & 0 & 0 \\
 2 & \frac{1}{2} & -\frac{1}{4} & 1 & 0 & -1 & 0 & 0 & 0 & 0 & 0 & 0 & 0 & 0 & 0 & 0 & 0 & 0 & 0 & 0 & 0 & 0 & 0 \\
 2 & \frac{1}{2} & -\frac{1}{4} & -1 & 0 & 1 & 0 & 0 & 0 & 0 & 0 & 0 & 0 & 0 & 0 & 0 & 0 & 0 & 0 & 0 & 0 & 0 & 0 \\
 2 & -\frac{1}{4} & \frac{1}{2} & 1 & 0 & 1 & 0 & 0 & 0 & 0 & 0 & 0 & 0 & 0 & 0 & 0 & 0 & 0 & 0 & 0 & 0 & 0 & 0 \\
 2 & -\frac{1}{4} & -\frac{1}{4} & 1 & 2 & 1 & 0 & 0 & 0 & 0 & 0 & 0 & 0 & 0 & 0 & 0 & 0 & 0 & 0 & 0 & 0 & 0 & 0 \\
 -1 & 0 & 0 & \frac{1}{2} & \frac{1}{2} & \frac{1}{2} & -\frac{1}{2} & 0 & -\frac{1}{2} & -\frac{1}{6} & -\frac{1}{6} & 0 & 0 & 0 & 0 & 0 & 0 & 0
& 0 & 0 & 0 & 0 & 0 \\
 -1 & 0 & 0 & 0 & \frac{1}{2} & \frac{1}{2} & \frac{1}{2} & -\frac{1}{2} & 0 & -\frac{1}{6} & \frac{1}{3} & 0 & 0 & 0 & 0 & 0 & 0 & 0 & 0 & 0 & 0
& 0 & 0 \\
 -1 & 0 & 0 & 0 & 0 & \frac{1}{2} & 0 & \frac{1}{2} & \frac{1}{2} & \frac{1}{3} & -\frac{1}{6} & 0 & 0 & 0 & 0 & 0 & 0 & 0 & 0 & 0 & 0 & 0 & 0 \\
 -1 & 0 & 0 & \frac{1}{2} & \frac{1}{2} & 0 & 0 & 0 & \frac{1}{2} & -\frac{1}{6} & \frac{1}{3} & 0 & 0 & 0 & 0 & 0 & 0 & 0 & 0 & 0 & 0 & 0 & 0 \\
 -1 & 0 & 0 & 0 & \frac{1}{2} & 0 & 0 & \frac{1}{2} & 0 & -\frac{1}{6} & -\frac{1}{6} & 0 & 0 & 0 & 0 & 0 & 0 & 0 & 0 & 0 & 0 & 0 & 0 \\
 -1 & 0 & 0 & 0 & 0 & -\frac{1}{2} & 0 & -\frac{1}{2} & -\frac{1}{2} & \frac{1}{3} & -\frac{1}{6} & 0 & 0 & 0 & 0 & 0 & 0 & 0 & 0 & 0 & 0 & 0 & 0
\\
 -1 & 0 & 0 & \frac{1}{2} & 0 & 0 & \frac{1}{2} & 0 & 0 & \frac{1}{3} & -\frac{1}{6} & 0 & 0 & 0 & 0 & 0 & 0 & 0 & 0 & 0 & 0 & 0 & 0 \\
 -1 & 0 & 0 & 0 & -\frac{1}{2} & 0 & 0 & -\frac{1}{2} & 0 & -\frac{1}{6} & -\frac{1}{6} & 0 & 0 & 0 & 0 & 0 & 0 & 0 & 0 & 0 & 0 & 0 & 0 \\
 -1 & 0 & 0 & 0 & -\frac{1}{2} & -\frac{1}{2} & -\frac{1}{2} & \frac{1}{2} & 0 & -\frac{1}{6} & \frac{1}{3} & 0 & 0 & 0 & 0 & 0 & 0 & 0 & 0 & 0 &
0 & 0 & 0 \\
 -1 & 0 & 0 & -\frac{1}{2} & 0 & 0 & -\frac{1}{2} & 0 & 0 & \frac{1}{3} & -\frac{1}{6} & 0 & 0 & 0 & 0 & 0 & 0 & 0 & 0 & 0 & 0 & 0 & 0 \\
 -1 & 0 & 0 & -\frac{1}{2} & -\frac{1}{2} & 0 & 0 & 0 & -\frac{1}{2} & -\frac{1}{6} & \frac{1}{3} & 0 & 0 & 0 & 0 & 0 & 0 & 0 & 0 & 0 & 0 & 0 & 0
\\
 -1 & 0 & 0 & -\frac{1}{2} & -\frac{1}{2} & -\frac{1}{2} & \frac{1}{2} & 0 & \frac{1}{2} & -\frac{1}{6} & -\frac{1}{6} & 0 & 0 & 0 & 0 & 0 & 0 &
0 & 0 & 0 & 0 & 0 & 0 \\
 1 & 0 & 0 & 0 & 0 & 0 & 0 & 0 & 0 & -\frac{1}{3} & \frac{1}{6} & 0 & 0 & 0 & 0 & 0 & 0 & 0 & 0 & 0 & 0 & 0 & 0 \\
 1 & 0 & 0 & 0 & 0 & 0 & 0 & 0 & 0 & \frac{1}{6} & -\frac{1}{3} & 0 & 0 & 0 & 0 & 0 & 0 & 0 & 0 & 0 & 0 & 0 & 0 \\
 1 & 0 & 0 & 0 & 0 & 0 & 0 & 0 & 0 & \frac{1}{6} & \frac{1}{6} & 0 & 0 & 0 & 0 & 0 & 0 & 0 & 0 & 0 & 0 & 0 & 0 \\
 0 & 0 & 0 & 0 & 0 & 0 & 0 & 0 & 0 & 0 & 0 & 0 & \frac{1}{3} & \frac{1}{3} & -\frac{1}{3} & -\frac{1}{2} & 0 & -\frac{32}{301} & -\frac{17}{110}
& -\frac{28}{27} & -\frac{8}{27} & 3 & 0 \\
 0 & 0 & 0 & 0 & 0 & 0 & 0 & 0 & 0 & 0 & 0 & \frac{1}{3} & 0 & \frac{1}{3} & -\frac{1}{3} & \frac{1}{2} & -\frac{1}{2} & -\frac{4}{301} & -\frac{3}{220}
& \frac{14}{9} & \frac{4}{9} & -\frac{1}{2} & \frac{7}{12} \\
 0 & 0 & 0 & 0 & 0 & 0 & 0 & 0 & 0 & 0 & 0 & \frac{1}{3} & \frac{1}{3} & 0 & -\frac{1}{3} & 0 & \frac{1}{2} & \frac{36}{301} & \frac{37}{220} & -\frac{14}{27}
& -\frac{4}{27} & -\frac{5}{2} & -\frac{7}{12} \\
 0 & 0 & 0 & 0 & 0 & 0 & 0 & 0 & 0 & 0 & 0 & \frac{1}{3} & -\frac{1}{3} & 0 & 0 & 1 & -\frac{1}{2} & -\frac{8}{43} & \frac{3}{220} & -\frac{14}{9}
& -\frac{4}{9} & -\frac{1}{2} & \frac{1}{4} \\
 0 & 0 & 0 & 0 & 0 & 0 & 0 & 0 & 0 & 0 & 0 & \frac{1}{3} & 0 & -\frac{1}{3} & 0 & \frac{1}{2} & \frac{1}{2} & \frac{24}{301} & -\frac{37}{220} &
\frac{14}{27} & \frac{4}{27} & \frac{7}{2} & -\frac{1}{4} \\
 0 & 0 & 0 & 0 & 0 & 0 & 0 & 0 & 0 & 0 & 0 & 0 & \frac{1}{3} & -\frac{1}{3} & 0 & -\frac{1}{2} & 1 & -\frac{60}{301} & 0 & 0 & 0 & -1 & \frac{5}{6}
\\
 0 & 0 & 0 & 0 & 0 & 0 & 0 & 0 & 0 & 0 & 0 & \frac{1}{3} & \frac{1}{3} & -1 & \frac{1}{6} & 0 & -\frac{1}{4} & -\frac{36}{301} & -\frac{8}{55} &
\frac{7}{27} & \frac{2}{27} & -\frac{3}{2} & -\frac{1}{3} \\
 0 & 0 & 0 & 0 & 0 & 0 & 0 & 0 & 0 & 0 & 0 & \frac{1}{3} & -1 & \frac{1}{3} & \frac{1}{6} & -\frac{1}{4} & \frac{1}{4} & \frac{52}{301} & -\frac{3}{110}
& \frac{1}{9} & -\frac{1}{9} & 0 & \frac{1}{3} \\
 0 & 0 & 0 & 0 & 0 & 0 & 0 & 0 & 0 & 0 & 0 & -1 & \frac{1}{3} & \frac{1}{3} & \frac{1}{6} & \frac{1}{4} & 0 & -\frac{16}{301} & \frac{19}{110} &
-\frac{10}{27} & \frac{1}{27} & \frac{3}{2} & 0 \\
 0 & 0 & 0 & 0 & 0 & 0 & 0 & 0 & 0 & 0 & 0 & \frac{1}{3} & 1 & -\frac{1}{3} & 0 & -\frac{1}{4} & -\frac{1}{4} & -\frac{16}{301} & \frac{19}{110}
& \frac{17}{27} & \frac{1}{27} & \frac{3}{2} & 0 \\
 0 & 0 & 0 & 0 & 0 & 0 & 0 & 0 & 0 & 0 & 0 & \frac{1}{3} & -\frac{1}{3} & 1 & 0 & -\frac{1}{2} & \frac{1}{4} & 0 & 0 & -1 & 0 & 0 & 0 \\
 0 & 0 & 0 & 0 & 0 & 0 & 0 & 0 & 0 & 0 & 0 & -1 & -\frac{1}{3} & -\frac{1}{3} & -\frac{1}{6} & -\frac{1}{4} & 0 & \frac{16}{301} & -\frac{19}{110}
& \frac{10}{27} & -\frac{1}{27} & -\frac{3}{2} & 0 \\
 0 & 0 & 0 & 0 & 0 & 0 & 0 & 0 & 0 & 0 & 0 & 1 & \frac{1}{3} & -\frac{1}{3} & 0 & \frac{1}{4} & -\frac{1}{2} & \frac{52}{301} & -\frac{3}{110} &
-\frac{8}{9} & -\frac{1}{9} & 0 & \frac{1}{3} \\
 0 & 0 & 0 & 0 & 0 & 0 & 0 & 0 & 0 & 0 & 0 & -\frac{1}{3} & \frac{1}{3} & 1 & 0 & \frac{1}{2} & -\frac{1}{4} & 0 & 0 & 1 & 0 & 0 & 0 \\
 0 & 0 & 0 & 0 & 0 & 0 & 0 & 0 & 0 & 0 & 0 & -\frac{1}{3} & -1 & -\frac{1}{3} & -\frac{1}{6} & \frac{1}{4} & -\frac{1}{4} & -\frac{52}{301} & \frac{3}{110}
& -\frac{1}{9} & \frac{1}{9} & 0 & -\frac{1}{3} \\
 0 & 0 & 0 & 0 & 0 & 0 & 0 & 0 & 0 & 0 & 0 & 1 & -\frac{1}{3} & \frac{1}{3} & 0 & -\frac{1}{4} & \frac{1}{2} & -\frac{52}{301} & \frac{3}{110} &
\frac{8}{9} & \frac{1}{9} & 0 & -\frac{1}{3} \\
 0 & 0 & 0 & 0 & 0 & 0 & 0 & 0 & 0 & 0 & 0 & -\frac{1}{3} & 1 & \frac{1}{3} & 0 & \frac{1}{4} & \frac{1}{4} & \frac{16}{301} & -\frac{19}{110} &
-\frac{17}{27} & -\frac{1}{27} & -\frac{3}{2} & 0 \\
 0 & 0 & 0 & 0 & 0 & 0 & 0 & 0 & 0 & 0 & 0 & -\frac{1}{3} & -\frac{1}{3} & -1 & -\frac{1}{6} & 0 & \frac{1}{4} & \frac{36}{301} & \frac{8}{55} &
-\frac{7}{27} & -\frac{2}{27} & \frac{3}{2} & \frac{1}{3} \\
 0 & 0 & 0 & 0 & 0 & 0 & 0 & 0 & 0 & 0 & 0 & 1 & 0 & 0 & 0 & 0 & 0 & 0 & 0 & 0 & 0 & 0 & 0 \\
 0 & 0 & 0 & 0 & 0 & 0 & 0 & 0 & 0 & 0 & 0 & 0 & 1 & 0 & 0 & 0 & 0 & 0 & 0 & 0 & 0 & 0 & 0 \\
 0 & 0 & 0 & 0 & 0 & 0 & 0 & 0 & 0 & 0 & 0 & 0 & 0 & 1 & 0 & 0 & 0 & 0 & 0 & 0 & 0 & 0 & 0 \\
 0 & 0 & 0 & 0 & 0 & 0 & 0 & 0 & 0 & 0 & 0 & 0 & 0 & 0 & 0 & 0 & 0 & 0 & 0 & 0 & 0 & 0 & 0 \\
 0 & 0 & 0 & 0 & 0 & 0 & 0 & 0 & 0 & 0 & 0 & 0 & 0 & 0 & 0 & 0 & 0 & 0 & 0 & 0 & 0 & 0 & 0 \\
 0 & 0 & 0 & 0 & 0 & 0 & 0 & 0 & 0 & 0 & 0 & 0 & 0 & 0 & 0 & 0 & 0 & 0 & 0 & 0 & 0 & 0 & 0 \\
 0 & 0 & 0 & 0 & 0 & 0 & 0 & 0 & 0 & 0 & 0 & 0 & 0 & 0 & 0 & 0 & 0 & 0 & 0 & 0 & 0 & 0 & 0 \\
 0 & 0 & 0 & 0 & 0 & 0 & 0 & 0 & 0 & 0 & 0 & 0 & 0 & 0 & 0 & 0 & 0 & 0 & 0 & 0 & 0 & 0 & 0 \\
 0 & 0 & 0 & 0 & 0 & 0 & 0 & 0 & 0 & 0 & 0 & 0 & 0 & 0 & 0 & 0 & 0 & 0 & 0 & 0 & 0 & 0 & 0 \\
 0 & 0 & 0 & 0 & 0 & 0 & 0 & 0 & 0 & 0 & 0 & 0 & 0 & 0 & 0 & 0 & 0 & 0 & 0 & 0 & 0 & 0 & 0 \\
 0 & 0 & 0 & 0 & 0 & 0 & 0 & 0 & 0 & 0 & 0 & 0 & 0 & 0 & 0 & 0 & 0 & 0 & 0 & 0 & 0 & 0 & 0 \\
\end{array}
\right)
}\nn
\ea

The $\ell_e$ vectors are defined by $(\ell_e)_t=\partial A^2_t/\partial L^2_e$. Using the basis (\ref{Eq:Basis}) the 10 vectors $\ell_e$, with $e$ running through the 4 principal axis edges $(e=\{0,1\},\{0,2\},\{0,4\},\{0,8\})$ and the 6 face diagonals $(e=\{0,3\},\{0,5\},\{0,6\},\{0,9\},\{0,10\},\{0,12\})$, are, in the limit $s\rightarrow 0$ and before decoupling of the spurious variables,  given by the columns of
\ba
M_{\ell}=
\frac{\lambda^{2}}{4\sqrt{2}}
{\scriptsize
\left(
\begin{array}{cccccccccc}
 1+\frac{1}{\omega _1} & 1+\frac{1}{\omega _2} & 0 & 0 & 0 & 0 & 0 & 0 & 0 & 0 \\
 1+\frac{1}{\omega _1} & 0 & 1+\frac{1}{\omega _4} & 0 & 0 & 0 & 0 & 0 & 0 & 0 \\
 1+\frac{1}{\omega _1} & 0 & 0 & 1+\frac{1}{\omega _8} & 0 & 0 & 0 & 0 & 0 & 0 \\
 0 & 1+\frac{1}{\omega _2} & 1+\frac{1}{\omega _4} & 0 & 0 & 0 & 0 & 0 & 0 & 0 \\
 0 & 1+\frac{1}{\omega _2} & 0 & 1+\frac{1}{\omega _8} & 0 & 0 & 0 & 0 & 0 & 0 \\
 0 & 0 & 1+\frac{1}{\omega _4} & 1+\frac{1}{\omega _8} & 0 & 0 & 0 & 0 & 0 & 0 \\
 2+\frac{2}{\omega _1} & 0 & 0 & 0 & 0 & 0 & 1+\frac{1}{\omega _2 \omega _4} & 0 & 0 & 0 \\
 2+\frac{2}{\omega _1} & 0 & 0 & 0 & 0 & 0 & 0 & 0 & 1+\frac{1}{\omega _2 \omega _8} & 0 \\
 2+\frac{2}{\omega _1} & 0 & 0 & 0 & 0 & 0 & 0 & 0 & 0 & 1+\frac{1}{\omega _4 \omega _8} \\
 0 & 2+\frac{2}{\omega _2} & 0 & 0 & 0 & 1+\frac{1}{\omega _1 \omega _4} & 0 & 0 & 0 & 0 \\
 0 & 2+\frac{2}{\omega _2} & 0 & 0 & 0 & 0 & 0 & 1+\frac{1}{\omega _1 \omega _8} & 0 & 0 \\
 0 & 2+\frac{2}{\omega _2} & 0 & 0 & 0 & 0 & 0 & 0 & 0 & 1+\frac{1}{\omega _4 \omega _8} \\
 0 & 0 & 2+\frac{2}{\omega _4} & 0 & 1+\frac{1}{\omega _1 \omega _2} & 0 & 0 & 0 & 0 & 0 \\
 0 & 0 & 2+\frac{2}{\omega _4} & 0 & 0 & 0 & 0 & 1+\frac{1}{\omega _1 \omega _8} & 0 & 0 \\
 0 & 0 & 2+\frac{2}{\omega _4} & 0 & 0 & 0 & 0 & 0 & 1+\frac{1}{\omega _2 \omega _8} & 0 \\
 0 & 0 & 0 & 2+\frac{2}{\omega _8} & 1+\frac{1}{\omega _1 \omega _2} & 0 & 0 & 0 & 0 & 0 \\
 0 & 0 & 0 & 2+\frac{2}{\omega _8} & 0 & 1+\frac{1}{\omega _1 \omega _4} & 0 & 0 & 0 & 0 \\
 0 & 0 & 0 & 2+\frac{2}{\omega _8} & 0 & 0 & 1+\frac{1}{\omega _2 \omega _4} & 0 & 0 & 0 \\
 0 & 0 & 0 & 0 & 2+\frac{2}{\omega _1 \omega _2} & 0 & 0 & 0 & 0 & 2+\frac{2}{\omega _4 \omega _8} \\
 0 & 0 & 0 & 0 & 0 & 2+\frac{2}{\omega _1 \omega _4} & 0 & 0 & 2+\frac{2}{\omega _2 \omega _8} & 0 \\
 0 & 0 & 0 & 0 & 0 & 0 & 2+\frac{2}{\omega _2 \omega _4} & 2+\frac{2}{\omega _1 \omega _8} & 0 & 0 \\
 -1+\frac{1}{\omega _1} & 1-\frac{1}{\omega _2} & 0 & 0 & 0 & 0 & 0 & 0 & 0 & 0 \\
 -1+\frac{1}{\omega _1} & 0 & 1-\frac{1}{\omega _4} & 0 & 0 & 0 & 0 & 0 & 0 & 0 \\
 -1+\frac{1}{\omega _1} & 0 & 0 & 1-\frac{1}{\omega _8} & 0 & 0 & 0 & 0 & 0 & 0 \\
 0 & -1+\frac{1}{\omega _2} & 1-\frac{1}{\omega _4} & 0 & 0 & 0 & 0 & 0 & 0 & 0 \\
 0 & -1+\frac{1}{\omega _2} & 0 & 1-\frac{1}{\omega _8} & 0 & 0 & 0 & 0 & 0 & 0 \\
 0 & 0 & -1+\frac{1}{\omega _4} & 1-\frac{1}{\omega _8} & 0 & 0 & 0 & 0 & 0 & 0 \\
 -2+\frac{2}{\omega _1} & 0 & 0 & 0 & 0 & 0 & 1-\frac{1}{\omega _2 \omega _4} & 0 & 0 & 0 \\
 -2+\frac{2}{\omega _1} & 0 & 0 & 0 & 0 & 0 & 0 & 0 & 1-\frac{1}{\omega _2 \omega _8} & 0 \\
 -2+\frac{2}{\omega _1} & 0 & 0 & 0 & 0 & 0 & 0 & 0 & 0 & 1-\frac{1}{\omega _4 \omega _8} \\
 0 & -2+\frac{2}{\omega _2} & 0 & 0 & 0 & 1-\frac{1}{\omega _1 \omega _4} & 0 & 0 & 0 & 0 \\
 0 & -2+\frac{2}{\omega _2} & 0 & 0 & 0 & 0 & 0 & 1-\frac{1}{\omega _1 \omega _8} & 0 & 0 \\
 0 & -2+\frac{2}{\omega _2} & 0 & 0 & 0 & 0 & 0 & 0 & 0 & 1-\frac{1}{\omega _4 \omega _8} \\
 0 & 0 & -2+\frac{2}{\omega _4} & 0 & 1-\frac{1}{\omega _1 \omega _2} & 0 & 0 & 0 & 0 & 0 \\
 0 & 0 & -2+\frac{2}{\omega _4} & 0 & 0 & 0 & 0 & 1-\frac{1}{\omega _1 \omega _8} & 0 & 0 \\
 0 & 0 & -2+\frac{2}{\omega _4} & 0 & 0 & 0 & 0 & 0 & 1-\frac{1}{\omega _2 \omega _8} & 0 \\
 0 & 0 & 0 & -2+\frac{2}{\omega _8} & 1-\frac{1}{\omega _1 \omega _2} & 0 & 0 & 0 & 0 & 0 \\
 0 & 0 & 0 & -2+\frac{2}{\omega _8} & 0 & 1-\frac{1}{\omega _1 \omega _4} & 0 & 0 & 0 & 0 \\
 0 & 0 & 0 & -2+\frac{2}{\omega _8} & 0 & 0 & 1-\frac{1}{\omega _2 \omega _4} & 0 & 0 & 0 \\
 0 & 0 & 0 & 0 & -2+\frac{2}{\omega _1 \omega _2} & 0 & 0 & 0 & 0 & 2-\frac{2}{\omega _4 \omega _8} \\
 0 & 0 & 0 & 0 & 0 & -2+\frac{2}{\omega _1 \omega _4} & 0 & 0 & 2-\frac{2}{\omega _2 \omega _8} & 0 \\
 0 & 0 & 0 & 0 & 0 & 0 & 2-\frac{2}{\omega _2 \omega _4} & -2+\frac{2}{\omega _1 \omega _8} & 0 & 0 \\
 3+\frac{3}{\omega _1} & 0 & 0 & 0 & 0 & 0 & 0 & 0 & 0 & 0 \\
 0 & 3+\frac{3}{\omega _2} & 0 & 0 & 0 & 0 & 0 & 0 & 0 & 0 \\
 0 & 0 & 3+\frac{3}{\omega _4} & 0 & 0 & 0 & 0 & 0 & 0 & 0 \\
 0 & 0 & 0 & 3+\frac{3}{\omega _8} & 0 & 0 & 0 & 0 & 0 & 0 \\
 -3+\frac{3}{\omega _1} & 0 & 0 & 0 & 0 & 0 & 0 & 0 & 0 & 0 \\
 0 & -3+\frac{3}{\omega _2} & 0 & 0 & 0 & 0 & 0 & 0 & 0 & 0 \\
 0 & 0 & -3+\frac{3}{\omega _4} & 0 & 0 & 0 & 0 & 0 & 0 & 0 \\
 0 & 0 & 0 & -3+\frac{3}{\omega _8} & 0 & 0 & 0 & 0 & 0 & 0 \\
\end{array}
\right)
} \; .
\ea

The four spurious length variables (corresponding to the body diagonals $e=\{0,7\},\{0,11\},\{0,13\},\{0,14\}$), in the limit $s\rightarrow 0$ and before decoupling, are encoded in
\ba
M_{\ell'}=
\frac{\lambda^{2}}{4\sqrt{2}}
{\scriptsize
\left(
\begin{array}{cccc}
 0 & 0 & 0 & 0 \\
 0 & 0 & 0 & 0 \\
 0 & 0 & 0 & 0 \\
 0 & 0 & 0 & 0 \\
 0 & 0 & 0 & 0 \\
 0 & 0 & 0 & 0 \\
 0 & 0 & 0 & 0 \\
 0 & 0 & 0 & 0 \\
 0 & 0 & 0 & 0 \\
 0 & 0 & 0 & 0 \\
 0 & 0 & 0 & 0 \\
 0 & 0 & 0 & 0 \\
 0 & 0 & 0 & 0 \\
 0 & 0 & 0 & 0 \\
 0 & 0 & 0 & 0 \\
 0 & 0 & 0 & 0 \\
 0 & 0 & 0 & 0 \\
 0 & 0 & 0 & 0 \\
 0 & 0 & 0 & 0 \\
 0 & 0 & 0 & 0 \\
 0 & 0 & 0 & 0 \\
 0 & 0 & 0 & 0 \\
 0 & 0 & 0 & 0 \\
 0 & 0 & 0 & 0 \\
 0 & 0 & 0 & 0 \\
 0 & 0 & 0 & 0 \\
 0 & 0 & 0 & 0 \\
 0 & 0 & 0 & 0 \\
 0 & 0 & 0 & 0 \\
 0 & 0 & 0 & 0 \\
 0 & 0 & 0 & 0 \\
 0 & 0 & 0 & 0 \\
 0 & 0 & 0 & 0 \\
 0 & 0 & 0 & 0 \\
 0 & 0 & 0 & 0 \\
 0 & 0 & 0 & 0 \\
 0 & 0 & 0 & 0 \\
 0 & 0 & 0 & 0 \\
 0 & 0 & 0 & 0 \\
 0 & 0 & 0 & 0 \\
 0 & 0 & 0 & 0 \\
 0 & 0 & 0 & 0 \\
 0 & 0 & 0 & 1+\frac{1}{\omega _2 \omega _4 \omega _8} \\
 0 & 0 & 1+\frac{1}{\omega _1 \omega _4 \omega _8} & 0 \\
 0 & 1+\frac{1}{\omega _1 \omega _2 \omega _8} & 0 & 0 \\
 1+\frac{1}{\omega _1 \omega _2 \omega _4} & 0 & 0 & 0 \\
 0 & 0 & 0 & 1-\frac{1}{\omega _2 \omega _4 \omega _8} \\
 0 & 0 & 1-\frac{1}{\omega _1 \omega _4 \omega _8} & 0 \\
 0 & 1-\frac{1}{\omega _1 \omega _2 \omega _8} & 0 & 0 \\
 1-\frac{1}{\omega _1 \omega _2 \omega _4} & 0 & 0 & 0 \\
\end{array}
\right)
} \q .
\ea

Next, we give a basis for the $\zeta$--sector: the basis consists of 13 vectors, 4 of which correspond to spurious variables. These appear as the last four columns of 
\ba
&&M_\zeta=\lambda^2 \times \nn\\
&&{\tiny
\left(
\begin{array}{ccccccccccccc}
 0 & 0 & 0 & 0 & 0 & 0 & 0 & 0 & 0 & 0 & 0 & 0 & 0 \\
 0 & 0 & 0 & 0 & 0 & 0 & 0 & 0 & 0 & 0 & 0 & 0 & 0 \\
 0 & 0 & 0 & 0 & 0 & 0 & 0 & 0 & 0 & 0 & 0 & 0 & 0 \\
 0 & 0 & 0 & 0 & 0 & 0 & 0 & 0 & 0 & 0 & 0 & 0 & 0 \\
 0 & 0 & 0 & 0 & 0 & 0 & 0 & 0 & 0 & 0 & 0 & 0 & 0 \\
 0 & 0 & 0 & 0 & 0 & 0 & 0 & 0 & 0 & 0 & 0 & 0 & 0 \\
 0 & 0 & 0 & 0 & 0 & 0 & 0 & 0 & 0 & 0 & 0 & 0 & 0 \\
 0 & 0 & 0 & 0 & 0 & 0 & 0 & 0 & 0 & 0 & 0 & 0 & 0 \\
 0 & 0 & 0 & 0 & 0 & 0 & 0 & 0 & 0 & 0 & 0 & 0 & 0 \\
 0 & 0 & 0 & 0 & 0 & 0 & 0 & 0 & 0 & 0 & 0 & 0 & 0 \\
 0 & 0 & 0 & 0 & 0 & 0 & 0 & 0 & 0 & 0 & 0 & 0 & 0 \\
 0 & 0 & 0 & 0 & 0 & 0 & 0 & 0 & 0 & 0 & 0 & 0 & 0 \\
 0 & 0 & 0 & 0 & 0 & 0 & 0 & 0 & 0 & 0 & 0 & 0 & 0 \\
 0 & 0 & 0 & 0 & 0 & 0 & 0 & 0 & 0 & 0 & 0 & 0 & 0 \\
 0 & 0 & 0 & 0 & 0 & 0 & 0 & 0 & 0 & 0 & 0 & 0 & 0 \\
 0 & 0 & 0 & 0 & 0 & 0 & 0 & 0 & 0 & 0 & 0 & 0 & 0 \\
 0 & 0 & 0 & 0 & 0 & 0 & 0 & 0 & 0 & 0 & 0 & 0 & 0 \\
 0 & 0 & 0 & 0 & 0 & 0 & 0 & 0 & 0 & 0 & 0 & 0 & 0 \\
 0 & 0 & 0 & 0 & 0 & 0 & 0 & 0 & 0 & 0 & 0 & 0 & 0 \\
 0 & 0 & 0 & 0 & 0 & 0 & 0 & 0 & 0 & 0 & 0 & 0 & 0 \\
 0 & 0 & 0 & 0 & 0 & 0 & 0 & 0 & 0 & 0 & 0 & 0 & 0 \\
 0 & 0 & -2 & 2 & 2 & -2 & 0 & 0 & 0 & 0 & 0 & 0 & 0 \\
 -2 & 2 & 0 & 0 & 2 & -2 & 0 & 0 & 0 & 0 & 0 & 0 & 0 \\
 -2 & 2 & -2 & 2 & 0 & 0 & 0 & 0 & 0 & 0 & 0 & 0 & 0 \\
 -2 & 2 & 2 & -2 & 0 & 0 & 0 & 0 & 0 & 0 & 0 & 0 & 0 \\
 -2 & 2 & 0 & 0 & -2 & 2 & 0 & 0 & 0 & 0 & 0 & 0 & 0 \\
 0 & 0 & -2 & 2 & -2 & 2 & 0 & 0 & 0 & 0 & 0 & 0 & 0 \\
 -2 & 2 & -2 & 2 & 4 \left(1+\sqrt{2}\right) & 4 \left(-1+\sqrt{2}\right) & 0 & 0 & 0 & 0 & 0 & 0 & 0 \\
 -2 & 2 & -4 & 4 & 2 & -2 & 0 & 4 \sqrt{2} & 0 & 0 & 0 & 0 & 0 \\
 -4 & 4 & -2 & 2 & 2 & -2 & 0 & 0 & 4 \sqrt{2} & 0 & 0 & 0 & 0 \\
 -2 & 2 & 4 \left(1+\sqrt{2}\right) & 4 \left(-1+\sqrt{2}\right) & -2 & 2 & 0 & 0 & 0 & 0 & 0 & 0 & 0 \\
 -2 & 2 & 2 & -2 & -4 & 4 & 4 \sqrt{2} & 0 & 0 & 0 & 0 & 0 & 0 \\
 -4 & 4 & 2 & -2 & -2 & 2 & 0 & 0 & 4 \sqrt{2} & 0 & 0 & 0 & 0 \\
 4 \left(1+\sqrt{2}\right) & 4 \left(-1+\sqrt{2}\right) & -2 & 2 & -2 & 2 & 0 & 0 & 0 & 0 & 0 & 0 & 0 \\
 2 & -2 & -2 & 2 & -4 & 4 & 4 \sqrt{2} & 0 & 0 & 0 & 0 & 0 & 0 \\
 2 & -2 & -4 & 4 & -2 & 2 & 0 & 4 \sqrt{2} & 0 & 0 & 0 & 0 & 0 \\
 4 \left(1+\sqrt{2}\right) & 4 \left(-1+\sqrt{2}\right) & 2 & -2 & 2 & -2 & 0 & 0 & 0 & 0 & 0 & 0 & 0 \\
 2 & -2 & 4 \left(1+\sqrt{2}\right) & 4 \left(-1+\sqrt{2}\right) & 2 & -2 & 0 & 0 & 0 & 0 & 0 & 0 & 0 \\
 2 & -2 & 2 & -2 & 4 \left(1+\sqrt{2}\right) & 4 \left(-1+\sqrt{2}\right) & 0 & 0 & 0 & 0 & 0 & 0 & 0 \\
 -8 \left(1+\sqrt{2}\right) & 8-8 \sqrt{2} & 0 & 0 & 0 & 0 & 0 & 0 & 8 \sqrt{2} & 0 & 0 & 0 & 0 \\
 0 & 0 & -8 \left(1+\sqrt{2}\right) & 8-8 \sqrt{2} & 0 & 0 & 0 & 8 \sqrt{2} & 0 & 0 & 0 & 0 & 0 \\
 0 & 0 & 0 & 0 & 8 \left(1+\sqrt{2}\right) & 8 \left(-1+\sqrt{2}\right) & -8 \sqrt{2} & 0 & 0 & 0 & 0 & 0 & 0 \\
 0 & 0 & 0 & 0 & 0 & 0 & 0 & 0 & 0 & 0 & 0 & 0 & 0 \\
 0 & 0 & 0 & 0 & 0 & 0 & 0 & 0 & 0 & 0 & 0 & 0 & 0 \\
 0 & 0 & 0 & 0 & 0 & 0 & 0 & 0 & 0 & 0 & 0 & 0 & 0 \\
 0 & 0 & 0 & 0 & 0 & 0 & 0 & 0 & 0 & 0 & 0 & 0 & 0 \\
 \frac{1}{12} \left(-45-2 \sqrt{2}\right) & \frac{1}{12} \left(45-2 \sqrt{2}\right) & \frac{1}{12} \left(-45-2 \sqrt{2}\right) & \frac{1}{12} \left(45-2
\sqrt{2}\right) & \frac{1}{12} \left(45+58 \sqrt{2}\right) & \frac{1}{12} \left(-45+58 \sqrt{2}\right) & -\frac{1}{3 \sqrt{2}} & \frac{29}{3 \sqrt{2}}
& \frac{29}{3 \sqrt{2}} & 1 & 1 & -2 & 0 \\
 \frac{1}{12} \left(-45-2 \sqrt{2}\right) & \frac{1}{12} \left(45-2 \sqrt{2}\right) & \frac{1}{12} \left(45+58 \sqrt{2}\right) & \frac{1}{12} \left(-45+58
\sqrt{2}\right) & \frac{1}{12} \left(-45-2 \sqrt{2}\right) & \frac{1}{12} \left(45-2 \sqrt{2}\right) & \frac{29}{3 \sqrt{2}} & -\frac{1}{3 \sqrt{2}}
& \frac{29}{3 \sqrt{2}} & 1 & 1 & 1 & -1 \\
 \frac{1}{12} \left(45+58 \sqrt{2}\right) & \frac{1}{12} \left(-45+58 \sqrt{2}\right) & \frac{1}{12} \left(-45-2 \sqrt{2}\right) & \frac{1}{12} \left(45-2
\sqrt{2}\right) & \frac{1}{12} \left(-45-2 \sqrt{2}\right) & \frac{1}{12} \left(45-2 \sqrt{2}\right) & \frac{29}{3 \sqrt{2}} & \frac{29}{3 \sqrt{2}}
& -\frac{1}{3 \sqrt{2}} & 1 & 1 & 1 & 1 \\
 \frac{1}{12} \left(45+58 \sqrt{2}\right) & \frac{1}{12} \left(-45+58 \sqrt{2}\right) & \frac{1}{12} \left(45+58 \sqrt{2}\right) & \frac{1}{12} \left(-45+58
\sqrt{2}\right) & \frac{1}{12} \left(45+58 \sqrt{2}\right) & \frac{1}{12} \left(-45+58 \sqrt{2}\right) & -\frac{1}{3 \sqrt{2}} & -\frac{1}{3 \sqrt{2}}
& -\frac{1}{3 \sqrt{2}} & -3 & 1 & 0 & 0 \\
\end{array}
\right)}.\;\q \nn
\ea

The columns of $M_\zeta$ give the vectors $\zeta_b$ in the $s\rightarrow 0$ limit and before decoupling of the four spurious variables. The basis defined in $M_\zeta$ leads to a particular simple form for  the $\Lambda^3$ part of the Hessian block $H_{(a)(b)}$.

\begin{acknowledgments}
  I thank  Seth Asante, Jos\'e  Padua-Arg\"uelles and Aldo Riello for discussions, as well as  Benjamin Knorr for interesting suggestions.
Research at Perimeter Institute is supported in part by the Government of Canada through the Department of Innovation, Science and Economic Development Canada and by the Province of Ontario through the Ministry of Colleges and Universities.
\end{acknowledgments}

\providecommand{\href}[2]{#2}
\begingroup
\endgroup

\end{document}